\documentstyle[seceq]{ptptex}

\def\msol{\ifmmode M_\odot\else$M_\odot$\fi}
\def\cosec{\rm cosec}

\notypesetlogo  

\title{
Self-Force in the Radiation Reaction Formula 
}
\subtitle{
Adiabatic Approximation of a Metric Perturbation and an Orbit
}    

\author{
Yasushi {\sc Mino}\footnote{E-mail: mino@tapir.caltech.edu}
}

\inst{
Mail code 130-33, California Institute of Technology, Pasadena, CA 91125 USA}

\recdate{
}

\abst{
We investigate a calculation method 
for the gravitational evolution of an extreme mass ratio binary, 
i.e. a binary constituting of a galactic black hole 
and a stellar mass compact object. 
The inspiralling stage of this system is considered 
to be a possible source of detectable gravitational waves. 
Because of the extreme mass ratio, 
one may approximate such a system by a black hole geometry 
(a Kerr black hole)
plus a linear metric perturbation induced by a point particle.
With this approximation, a self-force calculation 
was proposed for a practical calculation of the orbital evolution, 
including the effect of the gravitational radiation reaction, 
which is now known as the MiSaTaQuWa self-force\cite{self}. 
In addition, a radiation reaction formula was proposed\cite{RR} 
as an extension of the well-known balance formula 
of Press and Teukolsky\cite{PT}. 
The radiation reaction formula provides us a convenient method 
to calculate an infinite time averaged loss of ``the constants of motion" 
(i.e. the orbital energy, the $z$ component of the angular momentum, 
and the Carter constant) 
through the gravitational radiation reaction, 
with which one may approximately calculate the orbital evolution. 
Because these methods are approximately equivalent, 
we investigate the consequence of the orbital evolution 
using the radiation reaction formula. 

To this time, we have used 
the so-called adiabatic approximation of the orbital evolution 
and considered a method to evaluate the MiSaTaQuWa self-force 
by use of a linear metric perturbation. 
In this approach, we point out that 
there is a theoretical question concerning the choice of the gauge condition 
in the calculation of the MiSaTaQuWa self-force. 
Because of this gauge ambiguity, 
there is a case in which 
the MiSaTaQuWa self-force might not predict the orbital evolution 
in a physically expected manner, 
and this forces us to calculate a waveform 
only in the so-called dephasing time. 
We discuss the reason that such an unexpected thing happens 
and find that it is primarily because we consider 
the linear metric perturbation separately 
from the orbital evolution due to the self-force. 

We propose a new metric perturbation scheme 
under a possible constraint of the gauge conditions 
in which we obtain a physically expected prediction 
of the orbital evolution caused by the MiSaTaQuWa self-force. 
In this new scheme of a metric perturbation, 
an adiabatic approximation is applied 
to both the metric perturbation and the orbit. 
As a result, we are able to predict the gravitational evolution of the system 
in the so-called radiation reaction time scale, 
which is longer than the dephasing time scale. 
However, for gravitational wave detection by LISA, 
this may still be insufficient. 
We further consider a gauge transformation 
in this new metric perturbation scheme, 
and find a special gauge condition 
with which we can calculate the gravitational waveform 
of a time scale long enough for gravitational wave detection by LISA. 
}

\begin{document}

\maketitle 

\section{Introduction} \label{sec:intro}

Constructing a method for the accurate calculation of gravitational waveforms 
is an essential step toward success in gravitational wave detection. 
Since the Einstein equation is nonlinear, 
it is difficult to derive 
fully analytic solutions of every gravitational wave source, 
and we therefore consider an approximation method 
for a specific astrophysical source. 

In this paper, we consider an extreme mass ratio binary system.
X-ray and a variety of other observations suggest that 
there may exist supermassive black holes 
($M\sim 10^6--10^8\msol$) at the centers of galaxies. 
A number of stars have been observed at the galactic centers, 
and they interact with each other gravitationally. 
As a result of this multi-body interaction, 
we may expect that 
some galactic black holes may capture 
a stellar mass compact object ($m\sim 1--100\msol$) 
deeply into its gravitational potential without tidal disruption. 
We refer to such a pair of objects as an extreme mass ratio binary system, 
and such systems are believed to be possible sources 
of gravitational waves detectable by LISA\cite{LISA}.
Due to its orbital energy and angular momentum, 
the stellar mass compact object inspirals 
around the black hole for a long time, 
referred to as the `inspiral stage' of the binary evolution. 
During this stage, the system has a huge quadrupole moment 
and emits detectable gravitational waves, 
while the system slowly loses its orbital energy and angular momentum 
through this emission of gravitational waves. 

We consider a method to compute the gravitational waveform 
from the inspiral stage for such an extreme mass ratio binary. 
Because the black hole dominates the gravitational potential 
due to its huge mass, 
we assume that 
a metric perturbation provides a good approximation 
to describe the gravitational evolution of the system. 
We employ a Kerr black hole as the background 
because of the black hole uniqueness theorem in general relativity. 
Since the size of the stellar mass compact object 
is much smaller than the background curvature, 
it can be treated as a point particle 
to induce a metric perturbation\cite{match}. 
In this picture, the evolution of the binary can be interpreted 
as the orbital evolution of the particle around the black hole. 

It has been asserted that, 
because we already have a convenient formula 
to calculate a linear metric perturbation in a Kerr background\cite{PT}, 
the remaining issue in this problem is to develop 
a method to compute the orbit of a point particle. 
The orbital evolution can be derived by calculating the self-force, 
and a general formula for the self-force 
employing a linear metric perturbation scheme has been derived\cite{match}. 
Because the linear metric perturbation 
induced by a point particle diverges along the orbit, 
the implementation for the self-force 
involves a regularization calculation. 
We have made good progress 
in obtaining an exact derivation of the self-force\cite{self} 
by implementing this regularization calculation. 
However, there remains a problem with this method 
when we consider 
the self-force of a general orbit in a Kerr background. 
A related method that employs the so-called radiation reaction formula 
has also been proposed\cite{RR}. 
With this method, one can approximately derive the self-force, 
although it only contains the radiative part of the self-force. 
However, this method does not involve 
a complicated regularization calculation, 
and the numerical method has been established 
even for a general orbit in a Kerr background. 

Although the calculation of the self-force seems 
to be a reasonable step for a gravitational waveform calculation, 
the self-force is entirely gauge dependent 
and there is a question of what physical content 
the self-force actually has. 
In fact, one can easily show that, 
with an extreme choice of the gauge condition, 
the self-force can be made to vanish over the entire time interval 
for which a usual metric perturbation scheme is valid 
in describing a gravitational evolution of the system. 
We discuss that, although this is mathematically allowed 
in the usual metric perturbation scheme, 
the resulting description is technically disadvantageous 
from an energetics point of view. 
If we adopt such a gauge condition, we may eventually have to consider 
the gravitational evolution of the system in a nonperturbative manner, 
and we must use numerical relativity 
to calculate the nonlinear metric perturbation, to our knowledge. 

Instead of numerical relativity, 
we consider the extension of the usual metric perturbation 
with a certain constraint on a gauge condition. 
A primary reason for such an unexpected consequence 
in the usual metric perturbation scheme is 
that the particle must move along a geodesic 
to the leading order of the perturbation. 
As a result, a linear metric perturbation cannot track 
gravitational evolution 
together with the orbital evolution affected by the self-force. 
We propose a new metric perturbation scheme 
in which the MiSaTaQuWa self-force does not vanish. 
In this approach, the orbital evolution caused by this self-force turns 
to be essential to calculate the gravitational waveform. 

The new metric perturbation scheme 
may be able to predict the gravitational evolution of the system 
over a time scale longer than 
that of the usual metric perturbation. 
However, this time scale may not be sufficient 
to calculate a gravitational waveform for LISA. 
For this reason, we further consider a gauge transformation along an orbit 
so as to extend the validity of this new metric perturbation scheme. 
We find a special gauge condition 
with which the self-force contains 
only the radiation reaction component of the system, 
and, in this gauge condition, 
the adiabatic approximation of the linear metric perturbation 
can predict the evolution 
over a time scale long enough for LISA. 

This paper is organized as follows. 
In Sec.\ref{sec:gg}, we consider 
the gauge ambiguity of the MiSaTaQuWa self-force 
in the usual metric perturbation scheme 
and clarify the reason why such a prolblem could arise 
in the nonlinear theory of radiation reaction. 
In Sec.\ref{sec:adi}, we propose a new metric perturbation scheme, 
with which one can treat the gravitational evolution of the system 
together with the orbital evolution. 
In Sec.\ref{sec:gauge}, we consider a further gauge transformation 
in the new metric perturbation scheme 
and propose the so-called radiation reaction gauge condition.
We summarize our result in Sec.\ref{sec:sum}, 
with emphasis on clarifying the validity of 
this new metric perturbation scheme.

\section{Gauge ambiguity in the self-force problem} \label{sec:gg}

Gravitational radiation reaction is a physically real phenomenon, 
since we can define a gauge invariant momentum flux of gravitational waves 
in an asymptotic flat region of a background metric\cite{isc}. 
However, we find that a self-force is entirely gauge dependent 
consistently with a usual metric perturbation scheme 
and that it can be made to vanish 
through a special gauge choice along an orbit 
over the entire time scale 
for which the usual metric perturbation scheme is valid.

It is well-known that the self-force can be made to vanish 
at an orbital point by making the appropriate gauge choice, 
but we conjectured this might not be a problem, 
because the self-force would have non-vanishing components 
if we take the average of the self-force over a long time. 
In Ref.\cite{RR}, we show that this is true 
if we take the average of the self-force 
over an infinite time\footnote{
It is important to note that this argument depends 
only on the properties of the orbital equation, 
and it does not depend on the metric perturbation scheme.}, 
and we prove that non-vanishing components are gauge invariant 
in agreement with the gravitational radiation reaction. 
However, in the usual metric perturbation scheme, 
one cannot consider an infinite time average of the self-force 
because it is valid only over a finite time interval. 

In Subsec.\ref{ssec:per}, we consider a self-force 
in the usual metric perturbation scheme, 
and, in this scheme, we find that the average of the self-force 
can be made to vanish by making a special gauge choice along the orbit. 
This suggests that the relation between 
the self-force and the gravitational radiation reaction 
is entirely gauge dependent, 
and that it is not trivial as we previously believed. 
As a result, the so-called adiabatic approximation of the orbit\cite{PT0,PT1} 
would yield a gauge dependent prediction for the orbit. 
Because we expect that the orbital evolution caused by the self-force 
can be observed by the modulation of gravitational waves, 
a physically reasonable prediction of the orbit must be gauge invariant. 
This leads to the questions of 
how the self-force is related to the gravitational radiation reaction 
and what gauge conditions we should use to calculate the self-force. 
We discuss these questions in Subsec.\ref{ssec:eng} 

\subsection{Metric perturbation} \label{ssec:per}

We assume that 
a regularization calculation for the self-force 
is formally possible by using a matched asymptotic expansion 
of the metric of the near-zone expansion 
and the metric of the far-zone expansion, 
as considered in Ref.\cite{match}. 
Because the resulting self-force is defined 
by regularizing the metric of the far-zone expansion, 
we consider only the metric of the far-zone expansion. 
The metric of the far-zone expansion consists of the sum 
of a regular vacuum background metric and its metric perturbation 
induced by a point particle moving in the background metric. 
We assume the point particle has 
an appropriate internal structure\cite{Di} 
so that the metric of the far-zone expansion 
matches the metric of the near-zone expansion. 

Suppose we use the usual metric perturbation scheme 
for the calculation of the metric of the far-zone expansion. 
Then the far-zone expansion is defined as a perturbation expansion 
with the small parameter $m/L$ 
where $m$ is the Schwarzschild radius of the particle 
and $L$ is the curvature scale of the background along the orbit. 
In the usual metric perturbation scheme, 
we expand the metric and the stress-energy tensor 
in term of this small parameter as 
\begin{eqnarray}
g_{\mu\nu} &=& g^{(bg)}_{\mu\nu}
+(m/L) h^{(1)}_{\mu\nu}+(m/L)^2 h^{(2)}_{\mu\nu}+\cdots \,, \\
T^{\mu\nu} &=& (m/L) T^{(1)\mu\nu}
+(m/L)^2 T^{(2)\mu\nu}+\cdots \,, \label{eq:fe}
\end{eqnarray}
where $g^{(bg)}_{\mu\nu}$ is the vacuum background metric. 
For a valid perturbation, we must have 
\begin{eqnarray}
&& O(1) > |(m/L) h^{(1)}_{\mu\nu}| > 
|(m/L)^2 h^{(2)}_{\mu\nu}| > \cdots \,, \\
&& O(1) > |(m/L) T^{(1)\mu\nu}| > 
|(m/L)^2 T^{(2)\mu\nu}| > \cdots \,. 
\end{eqnarray}
We also expand the Einstein equation in $m/L$, 
and schematically we have 
\begin{eqnarray}
&& G^{(1)\mu\nu}[h^{(1)}] 
\,=\, T^{(1)\mu\nu} \,, \label{eq:ein1} \\
&& G^{(1)\mu\nu}[h^{(2)}]
+G^{(2)\mu\nu}[h^{(1)},h^{(1)}]
\,=\, T^{(2)\mu\nu} \,, \label{eq:ein2} \\
&& \cdots \,, \nonumber
\end{eqnarray}
where $G^{(1)\mu\nu}[h]$ and $G^{(2)\mu\nu}[h,h]$ 
are respectively the terms linear and quadratic in $h_{\mu\nu}$ 
of the Einstein tensor $G^{\mu\nu}[g+h]$. 

In Ref.\cite{match}, it is found that 
$T^{(1)\mu\nu}$ is the usual stress-energy tensor of a monopole particle 
for a matched asymptotic expansion of the metric. 
It is important to note that 
the linearized Einstein equation is algebraically divergence free 
with respect to the background metric, 
i.e. $G^{(1)\mu\nu}{}_{;\nu}=T^{(1)\mu\nu}{}_{;\nu}=0$. 
Thus, in this metric perturbation scheme, 
the particle must move along a geodesic 
in the background metric $g^{(bg)}_{\mu\nu}$ 
for a consistent solution of (\ref{eq:ein1}). 
The explicit form of $T^{(2)\mu\nu}$ can be derived 
by carrying out a further matched asymptotic expansion of the metric, 
but this has not been obtained yet. 
However, at least, it is clear that 
it has a term of the monopole particle 
coming from the deviation of the orbit from a geodesic 
because the MiSaTaQuWa self-force can be derived 
from LHS of (\ref{eq:ein2}) through a mass renormalization\cite{match}. 
This shows that 
in the expansion of the stress-energy tensor (\ref{eq:fe}), 
we assume an expansion of the orbit by $m/L$ as 
\begin{eqnarray}
z^\mu(\tau) &=& z^{(bg)\mu}(\tau)
+(m/L) z^{(1)\mu}+\cdots \,,\label{eq:oe}
\end{eqnarray}
where $\tau$ is defined to be the proper time in the background metric, 
and $z^{(bg)\mu}$ is a geodesic of the background metric 
given by $T^{(1)\mu\nu}{}_{;\nu}=0$. 
For a valid perturbation, we must have 
\begin{eqnarray}
&& O(1) > |(m/L) z^{(1)\mu}| > \cdots \,. \label{eq:ov} 
\end{eqnarray}
Intuitively, the orbit deviates from a geodesic 
as a result of the gravitational radiation reaction, 
and eventually the condition (\ref{eq:ov}) will be violated. 
Then, beyond this time, this perturbation scheme would fail 
to approximate the system. 

We consider a gauge transformation over a time interval 
during which the metric perturbation still approximates the system. 
The gauge transformation is defined as the small coordinate transformation 
$x^\mu \to \bar x^\mu = x^\mu+(m/L)\xi^\mu(x)$. 
The orbit transforms as 
$ z^\mu(\tau) \to \bar z^\mu(\tau)
= z^\mu(\tau+(m/L)\delta\tau)+(m/L)\xi^\mu(z(\tau))$,
where $\delta\tau$ is a function of $\tau$, 
so that $\tau$ remains the proper time 
of the new orbit $\bar z^\mu(\tau)$ in the background metric. 
Applying this to the perturbation expansion of the orbit (\ref{eq:oe}), 
we obtain the gauge transformation of $(m/L)z^{(1)\mu}$ as 
\begin{eqnarray}
(m/L)z^{(1)\mu}(\tau) &\to& (m/L)\bar z^{(1)\mu}(\tau) \nonumber \\ 
&& =\, (m/L)z^{(1)\mu}(\tau)
+(m/L)v^{(bg)\mu} \delta\tau+(m/L)\xi^\mu(z^{(bg)}(\tau)) 
\,, \label{eq:oo} 
\end{eqnarray}
where $v^{(bg)\mu}=dz^{(bg)}/d\tau$. 
We can eliminate $(m/L) z^{(1)\mu}(\tau)$ 
if the gauge transformation along the orbit satisfies 
\begin{eqnarray}
(m/L)\xi^\mu(z^{(bg)}(\tau)) &=& 
-(m/L)z^{(1)\mu}(\tau)+v^{(bg)\mu}(m/L)\Delta\tau \,, 
\end{eqnarray}
with an arbitrary function $\Delta\tau$. 
There always exists a gauge transformation 
which satisfies this condition over the entire time interval 
during which the metric perturbation scheme is valid, 
because it is smaller than $O(1)$ in accordance with (\ref{eq:ov}), 
and the orbit becomes a geodesic of the background metric 
under this gauge condition. 

Because the MiSaTaQuWa self-force causes a self-acceleration 
of the leading order in $m/L$-expansion 
and causes the orbit to deviate from the geodesic, 
the relation $(m/L)z^{(1)\mu}(\tau)=0$ implies 
that the MiSaTaQuWa self-force entirely 
vanishes under this gauge condition over the entire time interval 
for which this metric perturbation scheme is valid. 
This extreme example suggests that 
the MiSaTaQuWa self-force is completely gauge dependent 
in this usual metric perturbation scheme, 
and, even the time average of the self-force over a long time scale 
could be gauge dependent in general, 
as long as we use this perturbation scheme. 

At this stage, we see the mathematical reason 
why the self-force may not include 
the effect of gravitational radiation reaction. 
Because this perturbation scheme only allows a small deviation
from the background geodesic, as expressed by (\ref{eq:ov}), 
one could always eliminate this deviation
by a gauge transformation, as in (\ref{eq:oo}).
We note that this problem cannot be solved 
by calculating a non-linear metric perturbation 
because the problem resides in the metric perturbation scheme itself. 
This shows that we must modify the metric perturbation scheme 
so that we can describe 
a non-perturbative orbital deviation from the background geodesic 
as $(m/L)z^{(1)\mu}(\tau)\sim O(1)$. 
If the orbit can deviate from the geodesic nonperturbatively, 
it cannot be eliminated by the gauge transformation. 

This extreme example suggests that 
the relation between the self-force 
and the gravitational radiation reaction is not trivial, 
as we believed before\cite{PT0,PT1}. 
By an adiabatic approximation of an orbit, 
we usually use the time averaged part of the self-force. 
If this is a component averaged over an infinite time, 
it is known to be gauge invariant, 
and we have a gauge invariant prediction for the orbit. 
However, because the metric perturbation is not valid 
over an infinite time scale, 
it is not valid to consider an infinite time average, 
and it seems reasonable to take a finite time average. 
Then the orbital prediction becomes gauge dependent. 
In this situation, if we still use a linear metric perturbation, 
the gravitational waveform at infinity would also be gauge dependent, 
although we would expect it to be gauge invariant 
because it is observable. 

In the next subsection, 
we study this example from a different point of view, 
using a theory of a nonlinear metric perturbation. 
We expect that this extreme example 
may reveal to us the nature of the self-force, 
and, consequently we will be able to understand why and how 
we should take a certain class of gauge conditions for the self-force. 

\subsection{Energetics of the orbit and radiation} \label{ssec:eng}

Some idea of why such an unexpected thing happens 
can be gained from Ref.\cite{RR}. 
In Sec.III of Ref.\cite{RR}, 
we argue that the orbital energy does not decrease monotonically 
as a result of the emission of gravitational waves. 
We conjecture that gravitational radiation has its own energy 
and that the self-force describes the interaction 
between the orbital energy and the radiation energy. 
In fact, one can define an effective stress-energy tensor 
for the orbit and gravitational radiation, as argued in Ref.\cite{match}. 
As in the previous subsection, 
we consider only the metric of the far-zone expansion, 
and we suppose that 
the metric consists of the sum of the vacuum background metric 
and its perturbation induced by a point particle 
with an appropriate internal structure\cite{Di} 
as $g_{\mu\nu}= g^{(bg)}_{\mu\nu}+h_{\mu\nu}$. 

The Einstein equation for the far-zone expansion can be written formally as 
$G^{(1)}_{\mu\nu}[h]+G^{(2+)}_{\mu\nu}[h]=T_{\mu\nu}$, 
where $G^{(1)}_{\mu\nu}$ and $G^{(2+)}_{\mu\nu}$ 
contain the linear terms and the rest of the Einstein tensor 
with respect to $h_{\mu\nu}$, respectively. 
Because $G^{(1)}_{\mu\nu}$ is algebraically divergence free, 
one can define a conserved stress-energy tensor 
in the background metric as 
\begin{eqnarray}
{\cal T}^{\mu\nu} &=& T^{\mu\nu}-G^{(2+)\mu\nu}[h]
\,. \label{eq:einx}
\end{eqnarray}
We consider that 
the first and second terms of (\ref{eq:einx}) represent 
the effective stress-energy tensors 
for the orbit and the gravitational radiation, respectively, 
and neither of these is conserved in the background metric by itself. 
Although the effective stress-energy tensor (\ref{eq:einx}) 
is defined in a nonlinear manner in a general curved background, 
it reduces to the well-known gauge invariant stress-energy tensor 
for gravitational waves 
in the so-called short wavelength approximation 
and in the weak perturbation limit\cite{isc}. 

We suppose that 
the background metric $g^{(bg)}_{\mu\nu}$ is a Kerr black hole 
and that it has a Killing vector $\xi_\mu$. 
Then we can define total, orbital and radiation energies 
in the background metric as
\begin{eqnarray}
E^{(tot)} &=& E^{(orb)}+E^{(rad)} \,,\nonumber \\ 
E^{(orb)} &=& \int d\Sigma_\mu \xi_\nu T^{\mu\nu} \,, \quad 
E^{(rad)} \,=\, -\int d\Sigma_\mu \xi_\nu G^{(2+)\mu\nu} 
\,, \label{eq:ene} 
\end{eqnarray}
where the surface integral is taken 
over the spacelike hypersurface 
bounded by the future horizon and the future null infinity 
of the background black hole metric, 
and we assume an appropriate regularization 
along the orbit of the particle, 
say, by a matched asymptotic expansion. 
We note that 
the radiation energy $E^{(rad)}$ is not necessarily positive. 

Integrating (\ref{eq:einx}) 
over a small world tube surface around the orbit, 
one can derive the MiSaTaQuWa self-force\cite{match}. 
This shows that the self-force describes the interaction 
between the orbital energy and the gravitational radiation energy. 
The gauge ambiguity of the self-force could be 
interpreted as an ambiguity in defining 
the orbital energy and the radiation energy of (\ref{eq:ene}), 
since $T^{\mu\nu}$ and $G^{(2+)\mu\nu}[h]$ are gauge dependent. 
In order to see this, 
we apply the metric perturbation scheme presented in Subsec.\ref{ssec:per} 
to (\ref{eq:ene}). This yields 
\begin{eqnarray}
E^{(orb)} &=& (m/L)E^{(orb)(1)}+(m/L)E^{(orb)(2)}+\cdots 
\,, \nonumber \\
(m/L)E^{(orb)(1)} &=& 
\int d\Sigma_\mu \xi_\nu (m/L)T^{(1)\mu\nu} \,, \quad \cdots 
\label{eq:enex1} \\ 
E^{(rad)} &=& (m/L)^2E^{(rad)(2)}+\cdots 
\,, \nonumber \\
(m/L)^2E^{(rad)(2)} &=& 
-\int d\Sigma_\mu \xi_\nu (m/L)^2 G^{(2)\mu\nu}[h^{(1)},h^{(1)}] 
\,, \quad \cdots \,. \label{eq:enex2} 
\end{eqnarray}
In the usual perturbation scheme, 
$(m/L)T^{(1)\mu\nu}$ becomes 
the conserved stress-energy tensor of a monopole particle 
moving along a background geodesic, 
and $E^{(orb)(1)}$ becomes a constant of motion. 
$(m/L)^2T^{(2)\mu\nu}$ has not yet been derived explicitly, 
and calculations of $(m/L)^2E^{(orb)(2)}$ and $(m/L)^2E^{(rad)(2)}$ 
would involve regularization provided by the matched asymptotic expansion. 
However, it includes a contribution from 
the orbital deviation of the monopole particle 
relative to the background geodesic described by $(m/L)T^{(1)\mu\nu}$, 
as we discussed in the previous section. 
A gauge transformation of this part is defined by 
${\cal L}_{(m/L)\xi}(m/L)T^{(1)\mu\nu}$, 
and we find that the orbital energy transforms under gauge transformations as 
\begin{eqnarray}
(m/L)^2E^{(orb)(2)} &\to& (m/L)^2\bar E^{(orb)(2)} 
\nonumber \\ 
&&=\, (m/L)^2E^{(orb)(2)}+(m/L)^2\delta E \,, 
\end{eqnarray}
where $\delta E$ is arbitrary. 
Thus, the orbital energy is gauge dependent 
in the usual metric perturbation scheme. 

By contrast, the balance formula 
considers the gauge invariant momentum flux of gravitational waves 
at the future null infinity and the future horizon 
of the background black hole geometry. 
Hence, it describes the radiation reaction 
to the total energy, $E^{(tot)}$, 
rather than that to the orbital energy, $E^{(orb)}$\footnote{
The radiation reaction in the balance formula is gauge invariant\cite{isc}, 
and one may expect that 
the total energy is also gauge invariant. 
However, an explicit computation to prove 
its gauge invariance is not trivial 
because the definition of the total energy involves 
a gauge dependent regularization calculation 
and a robust volume averaging operation.}. 

Although we have not yet proposed a new metric perturbation scheme 
in which a metric perturbation would be valid 
over a much longer time scale (see the next section), 
the orbital energy and the radiation energy 
are defined in a nonlinear manner as (\ref{eq:ene}), 
and we cany consider what happens 
when we keep the orbital energy constant under a gauge transformation 
while the total energy continues to decrease 
through the gravitational radiation reaction 
over a sufficiently long time. 
Because the radiation energy, $E^{(rad)}$, is not necessarily positive, 
it is easily seen that 
the amplitude of the metric perturbation would inevitably increase, 
because the radiation energy, $E^{(rad)}$, would decrease 
as a result of the radiation reaction. 
Although this is mathematically allowed, 
it creates two technical problems: 
1) A linear approximation can no longer be used in calculating 
the evolution of the system gravitationally. 
2) The background coordinate system cannot be used 
as an approximate reference to measure the orbit. 

The above considerations reveal what is happening 
with regard to the behavior discussed in the previous subsection. 
It is reasonable to use a finite time averaged self-force 
with an arbitrary gauge condition 
for the adiabatic evolution of the orbit, 
and we would then have a gauge dependent prediction for the orbit. 
However, it is not reasonable to use a linear metric perturbation 
to calculate the gravitational waveform at infinity 
if we do not make an appropriate choice of the gauge condition 
for the self-force. 
We believe that 
the gauge invariant nature of the gravitational waveform at infinity 
can be recovered with an appropriate nonlinear calculation 
for the metric in this case. 

In order to avoid the difficulty involved in dealing with 
a nonlinear metric perturbation, 
it is disadvantageous to use such a gauge condition, 
and we therefore assume consider that 
the orbital energy must decrease through the radiation reaction 
for a reasonable choice of gauge conditions, 
so that one can understand the orbit 
in the background coordinate system as an approximate reference. 
We define the averaged energetic gauge condition along the orbit by 
\begin{eqnarray}
\left<{d\over dt}E^{(orb)}\right> &=& 
\left<{d\over dt}E^{(tot)}\right> \,, 
\label{eq:eg} 
\end{eqnarray}
where $<>$ represents a long-time average. 

Although the above condition is a necessary condition 
for a reasonable gauge choice, 
this may not be a sufficient condition, 
because the orbit in a Kerr background 
is not characterized by the energy $E^{(orb)}$ alone. 
In order to confirm that 
a certain gauge condition is actually advantageous, 
it is necessary to demonstrate that 
we can actually construct a new metric perturbation scheme 
that is valid over a longer time scale. 
For this, the new metric perturbation scheme 
must allow a nonperturbative deviation of an orbit from a geodesic, 
and therefore it would not be constrained by (\ref{eq:ov}). 
In the next section, we propose 
an adiabatic approximation of the metric perturbation 
in which the usual metric perturbation scheme is extended adiabatically.

\section{Adiabatic approximation} \label{sec:adi}

The adiabatic approximation of the orbital evolution 
is well known and is used in Ref.\cite{PT0,PT1}. 
It consists of the following two steps:\\
{\it 1) From the fact that the self-force is weak, 
we can instantaneously approximate the orbit by a geodesic, 
and, instead of the orbital coordinates, we consider 
the evolution of geodesic``constants''.}\\
{\it 2) We calculate the evolution of the geodesic ``constants'' 
caused by the gauge invariant momentum flux 
of gravitational waves induced by the approximated geodesic.}\\
In a Kerr background, Step 2) is insufficient 
because we do not have 
a gauge invariant radiation reaction to the Carter ``constant''. 
Instead of Step 2), 
the radiation reaction formula\cite{RR} includes the following step:\\
{\it 2') We calculate the evolution of geodesic ``constants'' 
employing the gauge invariant ``infinite time averaged'' self-force 
induced by the approximated geodesic.}\\
As shown in Ref.\cite{gal}, 
the ``infinite time averaged'' self-force 
acting on the orbital energy and the angular momentum 
is identical to the gauge invariant momentum flux of gravitational waves. 
Hence, 2') appears to be a generalization of 2). 
2') seems more reasonable than 2) because 2') is based on the self-force. 
However, one cannot take an infinite time average 
in the usual metric perturbation scheme 
because the metric perturbation scheme is valid only in an finite time. 
Instead of 2'), we can consider the following step:\\
{\it 2'') We calculate the evolution of geodesic ``constants'' 
employing the self-force induced by the approximated geodesic.}\\ 
In this case, the self-force is entirely gauge dependent, 
and the orbital evolution becomes gauge dependent 
as argued in the previous section. 

Because general relativity allows an arbitrary coordinate system, 
an orbit is always ``gauge dependent''. 
However, in a metric perturbation scheme, 
we use a fixed coordinate system of the background geometry, 
and we can define a gauge invariant object 
using the background metric as an approximate reference. 
To have a gauge dependent prediction for the orbit 
resulting from the self-force 
means that the coordinate system of the background geometry 
no longer acts as a proper reference for the full geometry. 
As shown in the previous section, 
this happens if the system cannot be described 
properly by a linear metric perturbation 
with an inappropriate gauge condition. 
This suggests that any argument for a self-force contains a gauge ambiguity 
until we consider the problem 
together with the evolution of the metric perturbation, 
so that we can carefully consider 
the validity of the metric perturbation during the orbital evolution. 
For this purpose, 
it is necessary to have a new metric perturbation scheme 
which predicts the orbital evolution consistently 
without the orbital constraint (\ref{eq:ov}). 

In Subsec.\ref{ssec:adm}, we propose 
an adiabatic approximation of the linear metric perturbation. 
In Subsec.\ref{ssec:adr}, we evaluate a metric perturbation 
in this new metric perturbation scheme 
based on the radiation reaction formalism\cite{RR}. 
In Subsec.\ref{ssec:ado}, we investigate the orbital evolution 
and consider the validity of this metric perturbation. 
The reason for using the radiation reaction formula is that 
we have an explicit general form of the self-force 
in the radiation reaction formula, 
and we can identify the component of the self-force 
which corresponds to 
the gauge invariant ``infinite time averaged'' momentum flux 
of gravitational waves\cite{RR}. (see (\ref{eq:rrf}))
Thus, one can easily formulate a new metric perturbation scheme 
under the averaged energetic gauge condition (\ref{eq:eg}) 
using this explicit form. 

\subsection{Adiabatic linear metric perturbation} \label{ssec:adm} 

As we summarize in Appendix \ref{app:geo}, 
an inspiralling geodesic in a Kerr geometry is conveniently characterized 
by $7$ constants with an orbital parameter $\lambda$. 
We  denote the orbital constants 
by $\gamma=\{{\cal E}^a;a=E,L,C;\lambda^b;b=r,\theta;C^c;c=t,\phi\}$. 
(See Appendix \ref{app:geo} for definitions.)
In the usual metric perturbation scheme, 
a linear metric perturbation is induced 
by a particle moving along a geodesic $\gamma$. 
Therefore, it is a function of $\gamma$, 
and we write $h^{(1)}_{\mu\nu}(x;\gamma)$. 

We consider extending this linear metric perturbation 
by including the effect of the self-force. 
Due to the self-force, the ``geodesic constants'' evolve 
as functions of the orbital parameter $\lambda$. 
We foliate the background spacetime into smooth spacelike hypersurfaces 
that intersect the orbit.\footnote{
The condition that the foliation surfaces are space-like is not important, 
as long as we consider the adiabatic approximation 
of the linear metric perturbation. 
For a well-defined nonlinear metric perturbation, 
there must be appropriate attenuation behavior 
of the adiabatic metric perturbation at the past null infinity. 
Otherwise, an asymptotic regularization is needed 
to calculate the nonlinear metric perturbation.\cite{amos}}
We define the foliation function $f(x)$ 
such that it is equal to the orbital parameter 
at the intersection with the orbit: $f(z(\lambda))=\lambda$. 
Using the foliation function, 
we define an adiabatic linear metric perturbation as 
\begin{eqnarray}
h^{ad(1)}_{\mu\nu}(x) &=& h^{(1)}_{\mu\nu}(x;\gamma(f(x))) \,. 
\label{eq:adm} 
\end{eqnarray}
Because there is an ambiguity in the foliation function, 
the adiabatic metric perturbation is not defined uniquely. 
However, our definition is sufficient 
for a leading correction to the gravitational evolution of the system. 

Because the derivative acts on $f(x)$, we have 
$[h^{(1)}_{\mu\nu;\lambda}(x;\gamma)]_{\gamma=\gamma(f(x))} 
\not = h^{ad(1)}_{\mu\nu;\lambda}(x)$. 
As a result, we cannot use the adiabatic linear metric perturbation 
as the metric of the far-zone expansion 
for the MiSaTaQuWa self-force in general.\cite{self} 
However, as long as $|(d/d\lambda)\gamma(\lambda)|$ is small, 
the result obtained with the MiSaTaQuWa self-force 
is approximately correct. 
We conjecture that the condition $|(d/d\lambda)\gamma(\lambda)|<O(1)$ 
is less severe than (\ref{eq:ov}) 
for the usual metric perturbation scheme 
and the adiabatic linear metric perturbation is valid 
in a longer time scale, as we discuss in the following subsections. 

An adiabatic linear metric perturbation does not satisfy 
the linearized Einstein equation (\ref{eq:ein1}). 
Hence, it is not defined 
by the usual metric perturbation scheme. 
Here we briefly discuss a systematic method to construct 
a post-adiabatic nonlinear metric perturbation. 
In addition to the small parameter used in the far-zone expansion $m/L$, 
we introduce another small parameter 
for the adiabatic expansion $\epsilon\sim |(d/d\lambda)\gamma|<O(1)$, 
and we expand the metric and the stress-energy tensor as 
\begin{eqnarray}
g_{\mu\nu} &=& g^{bg}_{\mu\nu} \nonumber \\ 
&& +(m/L)h^{(1,0)}_{\mu\nu}+(m/L)\epsilon h^{(1,1)}_{\mu\nu}
+(m/L)\epsilon^2 h^{(1,2)}_{\mu\nu}+\cdots \nonumber \\
&& +(m/L)^2h^{(2,0)}_{\mu\nu}+(m/L)^2\epsilon h^{(2,1)}_{\mu\nu}
+(m/L)^2\epsilon^2 h^{(2,2)}_{\mu\nu}+\cdots \nonumber \\
&& +\cdots \,, \\ 
T^{\mu\nu} &=& 
(m/L)T^{(1,0)\mu\nu}+(m/L)\epsilon T^{(1,1)\mu\nu}
+(m/L)\epsilon^2 T^{(1,2)\mu\nu}+\cdots \nonumber \\
&& +(m/L)^2T^{(2,0)\mu\nu}+(m/L)^2\epsilon T^{(2,1)\mu\nu}
+(m/L)^2\epsilon^2 T^{(2,2)\mu\nu}+\cdots \nonumber \\
&& +\cdots \,, 
\end{eqnarray}
In the limiting case $\epsilon\to 0$, 
this metric perturbation scheme becomes 
the usual metric perturbation scheme, 
and therefore some results for the usual metric perturbation scheme 
may apply to $h^{(n,0)}_{\mu\nu}$ and $T^{(n,0)\mu\nu}$. 
We define $h^{(1,0)}_{\mu\nu}$ to be $h^{ad(1)}_{\mu\nu}$ 
and $T^{(1,0)\mu\nu}$ to be the stress-energy tensor of a monopole particle. 
In contrast to the usual metric perturbation scheme, 
$T^{(1,0)\mu\nu}$ does not necessarily satisfy 
the conservation law in the background metric by itself, 
because the violation of the conservation law for $T^{(1,0)\mu\nu}$ 
can be compensated for by a higher order expansion 
with respect to $\epsilon$. 
Hence, the orbit of the particle described by $T^{(1,0)\mu\nu}$ 
is not necessarily a geodesic of the background, 
and the expansion of the orbit in the form (\ref{eq:oe}) 
is not necessary in this scheme. 

The violation of the conservation law presents a difficulty 
with regard to deriving equations 
for post-adiabatic linear metric perturbations, 
i.e., $h^{(0,m)}_{\mu\nu}$ with $m\not =1$. 
Schematically, an adiabatic linear metric perturbation 
satisfies the equation 
\begin{eqnarray}
G^{(1)\mu\nu}[(m/L)h^{ad(1)}] &=& (m/L)T^{(1,0)\mu\nu}
+\epsilon\Lambda^{(1,0)\mu\nu}[(m/L)h^{ad(1)}] \,, \label{eq:aein1}
\end{eqnarray}
where $\Lambda^{(1,1)\mu\nu}$ appears 
because $T^{(1,0)\mu\nu}$ does not satisfy the conservation law, 
while the LHS of (\ref{eq:aein1}) is algebraically divergence free. 
One may expect that the relation 
$G^{(1)\mu\nu}[(m/L)\epsilon h^{(1,1)}] =
(m/L)\epsilon T^{(1,1)\mu\nu}
-(m/L)\epsilon\Lambda^{(1,0)\mu\nu}[(m/L) h^{ad(1)}]$ 
holds for the first post-adiabatic linear metric perturbation $h^{(1,1)}$, 
where $(m/L)\epsilon T^{(1,1)\mu\nu}$ can be derived appropriately 
with a matched asymptotic expansion. 
However, this equation might be inconsistent, 
because the RHS is not necessarily divergence free, 
while the LHS is algebraically divergence free. 
We suppose that, as in the case of (\ref{eq:aein1}), 
there is a rank-two symmetric tensor 
that is $O((m/L)\epsilon^2)$ 
and cancels the divergence of the RHS. 
As a result, we obtain the following equations 
for the post-adiabatic linear metric perturbations: 
\begin{eqnarray}
&& G^{(1)\mu\nu}[(m/L)\epsilon^n h^{(1,n)}] \,=\, 
(m/L)\epsilon^n T^{(1,n)\mu\nu} 
\nonumber \\ && \qquad 
-\epsilon\Lambda^{(1,n-1)\mu\nu}[(m/L)\epsilon^{n-1}h^{(1,n-1)}] 
+\epsilon\Lambda^{(1,n)\mu\nu}[(m/L)\epsilon^n h^{(1,n)}] \,. 
\label{eq:aein2}
\end{eqnarray}
We believe that equations for nonlinear metric perturbations 
would be derived in a similar way. 

\subsection{Adiabatic extension of the radiation reaction formula} 
\label{ssec:adr} 

In order to see that 
the new metric perturbation scheme is valid 
over a longer time scale than the usual metric perturbation scheme, 
we extend the argument of the radiation reaction formula.\cite{RR} 
In the radiation reaction formula, 
a component of a self-force averaged over an infinite time 
is derived in the usual metric perturbation scheme 
from the general form of the self-force. 
It is shown that the loss of 
orbital energy and angular momentum due to the self-force 
averaged over an infinite amount of time 
is equal to that by the momentum flux 
caused by gravitational waves.\cite{gal} 
Since the self-force is entirely gauge dependent 
in the usual metric perturbation scheme, 
we can choose a gauge condition 
such that the rest of the self-force is small. 
Then we can say that the self-force of the radiation reaction formula 
is defined by the averaged energetic gauge condition (\ref{eq:eg}) 
over a time scale for which the usual metric perturbation scheme is 
valid.\footnote{A more complete argument for this gauge condition 
is given in Sec.\ref{sec:gauge} (see (\ref{eq:sfrr})).} 
We consider application of the adiabatic extension 
of the linear metric perturbation (\ref{eq:adm}) in this gauge condition. 
Because of the averaged energetic gauge condition (\ref{eq:eg}),
the radiation energy (\ref{eq:ene}) does not decrease by radiation reaction, 
and therefore we expect that the adiabatic metric perturbation scheme 
is valid over a time scale longer than 
that for the usual metric perturbation scheme. 

Using the result given in Appendix \ref{app:geo}, 
the stress-energy tensor of a geodesic 
$\gamma=\{{\cal E}^a,\lambda^b,C^c\}$ 
can be defined as 
\begin{eqnarray}
T^{\alpha\beta}_{(\gamma)}(x) &=& \int d\lambda 
s^{\alpha\beta}(\lambda)
\delta(t-z^t)\delta(r-z^r)\delta(\theta-z^\theta)\delta(\phi-z^\phi) 
\,, \\ 
s^{\alpha\beta}(\lambda) &=& \sum_{n_r,n_\theta} 
s^{\alpha\beta(n_r,n_\theta)}\exp[in_r\chi_r+in_\theta\chi_\theta] \,, 
\end{eqnarray}
where the expansion coefficients $s^{\alpha\beta(n_r,n_\theta)}$ 
are functions of ${\cal E}^a$. 
Because a Kerr black hole is a stationary and axisymmetric solution, 
we can assume that 
a Green function for a linear metric perturbation 
in the Boyer-Lindquist coordinates can be defined as 
\begin{eqnarray}
G_{\mu\nu\,\alpha'\beta'}(x,x') &=& 
\sum_{\omega,m}g_{\mu\nu\,\alpha'\beta'}^{(\omega,m)}(r,\theta;r',\theta')
\exp[-i\omega (t-t')+im(\phi-\phi')] \,. \label{eq:gm} 
\end{eqnarray}
The radiation reaction formula 
uses a linear metric perturbation derived with the Green method.\footnote{
Because the original radiation reaction formula uses 
a symmetry property of the linear metric perturbation itself 
rather than the metric perturbation 
derived with the Green function (\ref{eq:gm}), 
we present a derivation of the self-force in Appendix \ref{app:sf}.} 
It is schematically written as 
\begin{eqnarray}
h_{\mu\nu}(x;\gamma) &=& \sum_{\omega,m,n_r,n_\theta}
k^{(\omega,m,n_r,n_\theta)}(\gamma)
\nonumber \\ && \times 
h_{\mu\nu}^{(\omega,m,n_r,n_\theta)}({\cal E}^a;r,\theta)
\exp[-i\omega t+im \phi] \,, \label{eq:lm} \\
k^{(\omega,m,n_r,n_\theta)}(\gamma) &=& \int d\lambda
\tilde k^{(\omega,m,n_r,n_\theta)}({\cal E}^a)
\exp[i\omega\kappa_t-im\kappa_\phi
-in_r\chi_r-in_\theta\chi_\theta] \nonumber \\
&=& {2\pi\over \left<\dot Z^t\right>}
\tilde k^{(\omega,m,n_r,n_\theta)}({\cal E}^a)
e^{i(\omega C^t-m C^\phi)}
e^{-i2\pi(n_r\tilde\Omega_r\lambda^r
+n_\theta\tilde\Omega_\theta\lambda^\theta)}
\nonumber \\ 
&&\times \delta(\omega-2\pi\Omega_{m,n_r,n_\theta}) 
\,, \\ 
\Omega_{m,n_r,n_\theta} &=& 
m\Omega_\phi+n_r\Omega_r+n_\theta\Omega_\theta \,, \\ 
\Omega_\phi &=& {\left<\dot Z^\phi\right>\over \left<\dot Z^t\right>} 
\,, \quad 
\Omega_r = {\tilde\Omega_r\over \left<\dot Z^t\right>} \,, \quad 
\Omega_\theta = {\tilde\Omega_\theta\over \left<\dot Z^t\right>} \,. 
\end{eqnarray}
Here, $\Omega_\phi$, $\Omega_r$ and $\Omega_\theta$ 
are principal frequencies of the geodesic $\gamma$, 
and they are observable in the form of a spectral information 
concerning gravitational waves at infinity. 

Although (\ref{eq:lm}) looks general, 
the derivation of the linear metric perturbation 
constrains the gauge condition, 
because one may still add a gauge mode 
which is not a function of $t-t'$ or $\phi-\phi'$ to the Green function. 
We call this a physically reasonable class of gauge conditions, 
because one can easily read the principal frequencies 
from the resulting metric perturbation (\ref{eq:lm}), 
and because it satisfies 
the averaged energetic gauge condition (\ref{eq:eg}). 

For the adiabatic approximation 
of the linear metric perturbation (\ref{eq:adm}), 
we regard $\gamma$ as dynamical variables. 
With the foliation function $f(x)$, we have 
\begin{eqnarray}
h^{ad(1)}_{\mu\nu}(x) &=& \sum_{\omega,m,n_r,n_\theta}
k^{(\omega,m,n_r,n_\theta)}{(\gamma(f(x)))}
\nonumber \\ && \times 
h_{\mu\nu}^{(\omega,m,n_r,n_\theta)}({\cal E}^a(f(x));r,\theta)
\exp[-i\omega t+im \phi] \,. \label{eq:am}
\end{eqnarray}
The adiabatic linear metric perturbation satisfies 
the equation (\ref{eq:aein1}). 
Using the explicit form (\ref{eq:am}) 
of the adiabatic linear metric perturbation, 
it is seen that $\Lambda^{(1,0)\mu\nu}[h^{ad(1)}]$ is linear or quadratic 
in $(d/df)\gamma(f)$ or linear in $(d^2/df^2)\gamma(f)$. 
We find that the adiabatic linear metric satisfies 
the Einstein equation to an accuracy of $O(m/L)$
when $|(d/d\lambda)\gamma|$ and $|(d^2/d\lambda^2)\gamma|$ are small. 
In the next subsection, 
we investigate the evolution of $\gamma$ 
due to the adiabatic linear metric perturbation 
and show that the assumption of the adiabaticity is consistent. 

\subsection{Adiabatic evolution of an orbit} \label{ssec:ado}

Here, we study the orbital evolution 
caused by the adiabatic metric perturbation (\ref{eq:am}). 
Under the assumption that 
$|(d/d\lambda)\gamma|$ and $|(d^2/d\lambda^2)\gamma|$ are small, 
(\ref{eq:am}) approximates the linear metric perturbation, 
and we can still use the general expression of the self-force 
given in Ref.\citen{RR} 
by simply changing $\gamma$ to $\gamma(\lambda)$. This yields 
\begin{eqnarray}
{d\over d\lambda}{\cal E}^a &=& \sum_{n_r,n_\theta} 
\dot{\cal E}^{a(n_r,n_\theta)}({\cal E}^a(\lambda))
\exp[in_r\tilde\chi_r+in_\theta\tilde\chi_\theta] \,, \label{eq:rrf} 
\end{eqnarray}
where $\tilde\chi_b=2\pi\tilde\Omega_b({\cal E}^a(\lambda))
(\lambda+\lambda^b(\lambda))$.
For the orbital evolution, 
Ref.\citen{RR} uses the usual metric perturbation scheme 
and treats the orbital evolution 
by expanding the orbit as (\ref{eq:oe}), 
which is constrained by (\ref{eq:ov}). 
In the framework of the adiabatic metric perturbation, 
we cannot use (\ref{eq:oe}). 
Therefore we must treat the orbit in a nonperturbative manner. 
With the idea of the adiabatic evolution of the orbit, 
we consider the evolution of the orbital ``constants'' 
instead of the evolution of the coordinates.

Although Ref.\citen{RR} uses 
the perturbative expansion of the orbit (\ref{eq:oe}), 
the extrapolation of its result suggests 
\begin{eqnarray}
&& {\cal E}^a-{\cal E}^a_0 \,\sim\, O((m/L)(t/L)) 
\,, \nonumber \\ 
&& \lambda^b-\lambda^b_0 \, \,\sim\, O((m/L)(t/L)^2) \,, \quad 
C^c-C^c_0 \,\sim\, O((m/L)(t/L)^2) \,, \label{eq:orb}
\end{eqnarray}
where $\{{\cal E}^a_0,\lambda^b_0,C^c_0\}$ are the initial values. 
Thus, the evolution of the primary ``constants'' ${\cal E}^a$ 
becomes nonperturbative 
after the radiation reaction time, $T_{rad} \sim O(L(m/L)^{-1})$, 
whereas the evolution of the secondary ``constants'' $(\lambda^b,C^c)$ 
becomes nonperturbative 
after the dephasing time, $T_{dep} \sim O(L(m/L)^{-1/2})(<T_{rad})$. 
In Appendix \ref{app:orb}, we derive 
evolution equations for the secondary ``constants'', 
assuming that the evolution of the primary ``constants'' is perturbative, 
and we find that the approximation (\ref{eq:orb}) 
obtained with the usual metric perturbation scheme is qualitatively correct. 

Equation (\ref{eq:orb}) shows that the assumption of the adiabaticity 
is constrainted by the conditions 
$|(d/d\lambda)(\lambda^b,C^c)|<O(1)$ 
and $|(d^2/d\lambda^2)(\lambda^b,C^c)|<O(1)$, 
and the adiabaticity holds on the radiation reaction time scale, 
where the primary ``constants'' still evolve perturbatively. 
Hence the adiabatic linear metric perturbation 
satisfies the Einstein equation to an accuracy of $O(m/L)$ 
over the radiation reaction time scale $t<T_{rad}$. 

In Appendix \ref{app:toy}, we integrate the evolution equations 
in the case that 
the orbit is either eccentric in the equatorial plane ($\theta=\pi/2$)
or circular in an inclined precessing plane ($r=const$). 
In either case, it is shown that the circular (equatorial) orbit 
remains circular (equatorial) 
under the influence of the self-force.\cite{rrb} 
In Appendix \ref{app:toy}, we find that
the self-force in the adiabatic linear metric perturbation scheme 
can be used to predict the orbit over the radiation reaction time scale. 
The dominant part of the orbit is described 
by only the zero mode of the self-force, $\dot{\cal E}^{a(0,0)}$, 
which is the gauge invariant component of the self-force 
in the usual metric perturbation scheme.\cite{RR} 
This is a reasonable result of the adiabatic metric perturbation scheme. 
Because the adiabatic linear metric perturbation (\ref{eq:am}) 
is a valid metric perturbation on this time scale, 
the background metric accurately approximates the full geometry, 
and the orbit using the background coordinates as an approximate reference 
must be gauge invariant to leading order. 
We also find that this is observable. 
Because the adiabatic metric perturbation scheme 
describes the metric perturbation 
together with the orbital evolution caused by the self-force, 
the modulation of gravitational waves induced by the radiation reaction 
can be read off of (\ref{eq:am}). 
In (\ref{eq:am}), we have 
\begin{eqnarray}
k^{(\omega,m,n_r,n_\theta)}(\gamma(f))
\sim e^{-i2\pi(n_r\tilde\Omega_r\lambda^r(f)
+n_\theta\tilde\Omega_\theta\lambda^\theta(f))}
e^{i(\omega C^t(f)-m C^\phi(f))} \,, 
\end{eqnarray}
and the nonperturbative evolution of $(\lambda^b,C^c)$ 
can be observed as a modulation of the wave phase. 

The remaining component of the orbit involves 
nonzero modes of the self-force (\ref{eq:rrf}), 
$\dot{\cal E}^{a(n_r,n_\theta)}$, with $(n_r,n_\theta)\not =(0,0)$. 
They impart an $O((m/L))$ contribution to the orbit 
on the radiation reaction time scale. 
This is the same order as the metric perturbation, 
and therefore these modes are interpreted as a gauge. 
In the next section, we show that 
the nonzero mode of the self-force 
$\dot{\cal E}^{a(n_r,n_\theta)}$, with $(n_r,n_\theta)\not =(0,0)$
is gauge dependent 
in the adiabatic metric perturbation scheme. 
We also find that 
the adiabatic nonlinear metric perturbation imparts 
an $O((m/L)^2(t/L)^2)$ contribution to the orbit. 
Well within the radiation reaction time scale, 
this is smaller than $O(1)$, 
and it is consistent with the fact 
that the adiabatic linear metric perturbation 
satisfies the Einstein equation to an accuracy of $O(m/L)$ 
on the radiation reaction time scale. 

Although the above may seem quite reasonable,
there is an additional concern when we consider a general orbit.
By looking at (\ref{eq:rrf}), one can easily understand the reason 
why the gauge invariant component of the self-force 
$\dot{\cal E}^{a(0,0)}$ imparts a dominant contribution to the orbit. 
In the adiabatic metric perturbation scheme, 
$\dot{\cal E}^{a(0,0)}$ slowly changes 
over the radiation reaction time scale 
and it coherently contributes to the orbital evolution. 
Contrastingly, 
the nonzero modes of the self-force, $(n_r,n_\theta)\not=(0,0)$ 
give a small contribution to the orbital evolution, 
because the phase $n_r\tilde\chi_r+n_\theta\tilde\chi_\theta$ 
changes on the dynamical time scale in general, 
and the contribution cancels over a longer time scale. 
This is true for a circular or equatorial orbit. 
However, there could be a case in which 
the phase $n_r\tilde\chi_r+n_\theta\tilde\chi_\theta$ of a nonzero mode 
becomes almost stationary. 

This problem appears as a discrepancy between 
the gauge invariant quantity in the radiation reaction formula 
and in the balance formula 
in the framework of the usual metric perturbation scheme. 
In Ref.\citen{RR}, the gauge invariant component of the self-force 
is derived by taking an infinite time average over the self-force. 
The infinite time averaged self-force is derived 
in the usual metric perturbation as 
$\left<{d\over d\lambda}{\cal E}^a\right> = 
\sum\dot{\cal E}^{a(n_r,n_\theta)}$, 
where the sum is taken over the values of $n_r$ and $n_\theta$ such that 
$n_r\tilde\Omega_r+n_\theta\tilde\Omega_\theta=0$.
On the other hand, Ref.\citen{gal} shows that 
the zero mode, $\dot{\cal E}^{a(0,0)}$, 
is equal to the gauge invariant momentum flux of gravitational waves. 
In order to resolve this disagreement, 
it is assumed in Ref.\citen{RR} that 
$\tilde\Omega_r/\tilde\Omega_\theta$ is irrational. 
In this case, the infinite time averaged component of the self-force 
becomes equal to the gauge invariant momentum flux 
of gravitational waves. 
However, this assumption is not strictly valid. 
If there is a nonzero mode for which 
$n_r\tilde\Omega_r+n_\theta\tilde\Omega_\theta=0$, 
the phase $n_r\tilde\chi_r+n_\theta\tilde\chi_\theta$ 
becomes almost stationary, 
and it may contribute to the orbital evolution. 
In this case, we may need to develop another formula, 
since we can only calculate $\dot{\cal E}^{a(0,0)}$ 
using the radiation reaction formula.\cite{RR} 

By extending the analysis of Appendic \ref{app:toy}, 
we conjecture that such a problem might not arise 
in the adiabatic metric perturbation scheme 
and that all we need for an orbital prediction is $\dot{\cal E}^{a(0,0)}$. 
The primary ``constants'' for a general orbit 
during the radiation reaction time, $t<O(L(m/L)^{-1})$, evolve according to 
\begin{eqnarray}
{\cal E}^a &=& {\cal E}^a_0 +\int^\lambda_{\lambda_0} d\lambda
\sum_{n_r,n_\theta} \dot{\cal E}^{a(n_r,n_\theta)}({\cal E}_0)
\exp[in_r\tilde\chi_r+in_\theta\tilde\chi_\theta] +O(\mu^2 t)
\,. \label{eq:ee}
\end{eqnarray}
During the dephasing time interval $0<t<O(L(m/L)^{-1/2})$,
we have $\tilde\chi_b\sim\tilde\Omega_b\lambda+const$. 
As a result, the phase of any nonzero mode with 
$|n_r\tilde\Omega_r+n_\theta\tilde\Omega_\theta|<O(1/L(m/L)^{1/2})$ 
becomes stationary 
and gives a coherent $O((m/L)^{1/2})$ contribution 
to the primary ``constants'' given in (\ref{eq:ee}). 
Beyond the dephasing time, 
$\tilde\chi_b$ begins to evolve in a nonperturbative manner, 
and, even if $|n_r\tilde\Omega_r+n_\theta\tilde\Omega_\theta|$ is small,
$n_r\tilde\chi_r+n_\theta\tilde\chi_\theta$ would not be stationary 
unless $(n_r,n_\theta)=(0,0)$. 
During the time interval satisfying $O(L(m/L)^{-1/2})<t<O(L(m/L)^{-2/3})$, 
a mode for which $|(n_r\tilde\Omega_{r,a}+n_\theta\tilde\Omega_{\theta,a})
\dot{\cal E}^{a(0,0)}|< O(1/L(m/L)^2)$
would have a stationary phase. 
Extending this approximation, 
during the time interval $O(L(m/L)^{-n/(n+1)})<t<O(L(m/L)^{-(n+1)/(n+2)})$, 
a mode for which $|(n_r\tilde\Omega_{r,a}+n_\theta\tilde\Omega_{\theta,a})
(\dot{\cal E}^{a(0,0)})^n|< O(1/L(m/L)^{n+1})$
would have a stationary phase, 
and it would contribute an $O((m/L)^{1/(n+2)})<O(1)$ term to (\ref{eq:ee}) 
during this time interval. 
Thus, only the zero mode $\dot{\cal E}^{a(0,0)}$ could impart 
a coherent contribution of $O(1)$ over the radiation reaction time scale. 

Before further investigating the gauge condition, 
we discuss the validity of the adiabatic metric perturbation 
for an application to the LISA project. 
The LISA project is planning to detect gravitational waves 
for several years, 
and we need theoretical templates of gravitational waveforms 
valid over such time intervals. 
It is estimated that the mass ratio of a possible target binary 
would be $\sim 10^{-5}$. 
Suppose $m/L=10^{-5}$ with a dynamical time scale $L=100s$. 
Then the radiation reaction time $T_{rad}$ would be 
on the order of several months. 
Hence, a theoretical prediction obtained 
using the adiabatic linear metric perturbation 
may not be reliable enough for the LISA project. 

In the next section, we argue that fixing the gauge condition 
improves the validity of an adiabatic linear metric perturbation, 
and with this improvement one can calculate gravitational waveforms 
with the adiabatic metric perturbation in a more reliable manner.

\section{Radiation reaction gauge}
\label{sec:gauge}

In this section, we consider a further gauge transformation 
of an adiabatic metric perturbation. 
It is found in Ref.\citen{RR} that 
in the usual metric perturbation scheme, 
the gauge transformation of the self-force 
acting on the primary ``constants'' 
can be written in terms of a total derivative along the orbit. 
For this reason, the gauge transformation of the primary ``constants'' 
can be expressed locally. 
Because the argument used here depends 
on only the properties of the orbital equation, 
the same result can be used for an adiabatic metric perturbation. 
However, for technical simplicity, 
here we first consider the gauge transformation 
in the usual metric perturbation scheme. 
Then we extend the result 
to the case of an adiabatic metric perturbation. 

Suppose we transform the coordinates as $x^\mu \to x^\mu +\xi^\mu$. 
Then, to leading order in the gauge vector $\xi^\mu$, 
we have the primary ``constants'' transform as 
\begin{eqnarray}
\delta {\cal E}^{E/L} \,=\, 
-\eta^{E/L}_\alpha v^\beta \xi^\alpha_{;\beta} 
+\eta^{E/L}_{\alpha;\beta} v^\beta \xi^\alpha \,,\quad 
\delta {\cal C}^C \,=\,
-\eta^C_{\alpha\beta} v^\beta v^\gamma \xi^\alpha_{;\gamma}
\,, \label{eq:ec_gau} 
\end{eqnarray}
where $v^\alpha$ is the $4$-vector of the orbit, 
$\eta^{E/L}_\alpha$ are the Killing vectors 
used to define the orbital energy 
and the $z$ component of the angular momentum, 
and $\eta^C_{\alpha\beta}$ is the Killing tensor 
used to define Carter constant. 
In the usual metric perturbation scheme, 
the primary ``constants'' evolve according to the relation 
\begin{eqnarray}
{\cal E}^a &=& {\cal E}^a_0 +\dot{\cal E}^{a(0,0)}\lambda
\nonumber \\ 
&&+\sum_{(n_r,n_\theta)\not=(0,0)} 
{1\over i(n_r\tilde\Omega_r+n_\theta\tilde\Omega_\theta)}
\dot{\cal E}^{a(n_r,n_\theta)}
\exp[i(n_r\tilde\Omega_r+in_\theta\tilde\Omega_\theta)\lambda] 
\,, \label{eq:pee} 
\end{eqnarray}
where we have excluded the exceptional case 
that $\tilde\Omega_r/\tilde\Omega_\theta$ is rational, 
as in Ref.\citen{RR}. 
We now consider the elimination of the last oscillating part of (\ref{eq:pee}) 
through a gauge transformation. 
Although $\xi^\alpha$ is a vector field defined in the entire spacetime, 
the following analysis considers 
the condition for $\xi^\alpha$ along the orbit. 
Therefore, we consider $\xi^\alpha$ to be a function of the orbital parameter 
as $\xi^\alpha=\xi^\alpha(z(\lambda))$. 

As is shown in Appendix \ref{app:gau},
the gauge equation (\ref{eq:ec_gau}) becomes 
\begin{eqnarray}
\delta {\cal E}^a &=& \sum_{b=E,L,C}\left(
A^a{}_b {d\over d\lambda}\xi^b +B^a{}_b \xi^b \right) \,,
\end{eqnarray}
where ${\bf A}$ is a regular $3\times 3$-symmetric matrix
and ${\bf B}$ is a $3\times 3$ anti-symmetric matrix.
With these properties,
one can formally integrate the gauge equation as
\begin{eqnarray}
{\bf \xi} &=& {\bf C}^{-1}
\int d\lambda {\bf C}{\bf A}^{-1}\delta{\cal E} 
\,, \label{eq:egg}
\end{eqnarray}
where ${\bf C}$ is a matrix schematically defined by
\begin{eqnarray}
{\bf C} &=& \exp(\int d\lambda {\bf A}^{-1}{\bf B}) \,.
\end{eqnarray}
We recall that,
because the components of the matrix ${\bf A}^{-1}{\bf B}$ 
are functions of $r(\lambda)$ and $\theta(\lambda)$,
their components can be Fourier-decomposed
as $\exp(in_r\chi_r+in_\theta\chi_\theta)$ (see Appendix \ref{app:geo}). 
Applying an appropriate mathematical transformation,
the matrix ${\bf C}$ becomes
\begin{eqnarray}
{\bf C} &=& \exp(ic_1{\bf u}_1+ic_2{\bf u}_2) \,,
\end{eqnarray}
where ${\bf u}_1$ and ${\bf u}_2$ are
two independent elements of the $so(3)$-algebra,
and two coefficients $c_1$ and $c_2$ are functions of $\lambda$ 
that satisfy
\begin{eqnarray}
c_{1/2} &=& \dot c_{1/2}  \lambda +\sum
c^{(n_r,n_\theta)}_{1/2}\exp(in_r\chi_r+in_\theta\chi_\theta) \,.
\end{eqnarray}
For this reason, the matrix ${\bf C}$ can be decomposed 
into discrete Fourier modes, and we schematically write
\begin{eqnarray}
{\bf C} &=& \sum_\omega {\bf C}_\omega \exp(i\omega \lambda) \,.
\end{eqnarray}

In order to eliminate the oscillating term of (\ref{eq:pee}),
we define $\delta {\cal E}^a$ for (\ref{eq:egg}) as 
\begin{eqnarray}
\delta {\cal E}^a &=& \sum_{n_r,n_\theta}
{\cal E}^{a(n_r,n_\theta)}\exp[in_r\chi_r+in_\theta\chi_\theta]
-\delta {\cal E}^a_0 \,,
\end{eqnarray}
where $\delta {\cal E}^a_0$ is a constant.
If we set $\delta {\cal E}^a_0=0$, 
we would have $\xi^\alpha \sim O((m/L)\lambda)$, 
due to the integration over $\lambda$ in (\ref{eq:egg}), 
and as a result, the gauge transformation of the metric perturbation 
would grow as $O((m/L)(t/L)) $ around the orbit. 
Such a gauge transformation would cause 
the metric perturbation to become invalid in a short time 
in the adiabatic metric perturbation scheme. 
In order to avoid this, 
we choose $\delta {\cal E}^a_0 \not=0$, 
so that the gauge transformation (\ref{eq:egg}) becomes oscillatory. 
Then we have 
\begin{eqnarray}
{\cal E}^a &=& ({\cal E}^a_0 +\delta {\cal E}^a_0)
+\dot{\cal E}^{a(0,0)}\lambda \,. \label{eq:eex}
\end{eqnarray}
This shows that nonzero modes, 
$\dot{\cal E}^{a(n_r,n_\theta)}$ with $(n_r,n_\theta)\not=(0,0)$, 
are completely gauge dependent. 
We recall that the initial value ${\cal E}^a_0$ 
has a $O(m/L)$ ambiguity 
when we include the metric perturbation. 
Therefore, $\delta {\cal E}^a_0$ can be renormalized 
into the initial value, ${\cal E}^a_0$. 

We now consider the self-force in the adiabatic metric perturbation scheme. 
The self-force under this gauge condition becomes 
\begin{eqnarray}
{d\over d\lambda}{\cal E}^a &=& \dot{\cal E}^{a(0,0)} 
\,. \label{eq:sfrr}
\end{eqnarray}
We refer to this gauge condition as the ``radiation reaction gauge'', 
as the self-force is written 
solely in terms of the gauge invariant dissipative term. 
It is interesting to note that 
a previous calculation employing the balance formula\cite{PT0,PT1} 
uses the same equation. 
However, it was considered to be an approximation 
and a validity of this approximation has been discussed.\cite{PT0} 
The derivation here uses the gauge freedom of the metric perturbation. 
Hence, it is an exact result 
in the framework of the adiabatic metric perturbation. 

The application of this result to a motion of a spinning particle 
is interesting. 
A spinning particle does not move along a geodesic 
due to the spin-curvature coupling,\cite{papa} 
and one may regard this effect as a force: 
\begin{eqnarray}
{D\over d\lambda}v^\alpha &=& F^\alpha_{spin} \,.
\end{eqnarray}
However, this is a conservative effect, 
and the conservative force can be eliminated 
by choosing the radiation reaction gauge, 
as long as the force is on the order of the metric perturbation, 
$F^\alpha_{spin} \sim O(m/L)$. 
This is the case for a spinning compact object, 
and the result suggests that 
we may not observe the spin effect by gravitational wave detection. 
Although the force resulting from the spin-curvature coupling 
can be eliminated, 
this does not mean that there is no such effect. 
As (\ref{eq:eex}) shows, 
it is simply absorbed into the initial value. 
As a result, it alters the condition 
for the last stable orbit of the binary. 

It is interesting that the radiation reaction gauge changes 
the validity argument for an adiabatic linear metric perturbation. 
Because $(d/d\lambda){\cal E}^a$ does not have nonzero modes, 
we have $(d^2/d\lambda^2){\cal E}^a\sim O((m/L)^2)$ 
under this gauge condition. 
We also have $((d^2/d\lambda^2)\lambda^b,(d^2/d\lambda^2)C^c)\sim O(m/L)$ 
and the validity of the adiabatic linear metric perturbation 
depends on only $((d/d\lambda)\lambda^b,(d/d\lambda)C^c)$. 

Here we use the post-Newtonian approximation 
for $(d/d\lambda){\cal E}^a=\dot{\cal E}^{a(0,0)}$, 
because $\dot{\cal E}^{a(0,0)}$ is 
a throughly studied gauge invariant quantity.\cite{PN} 
Although the post-Newtonian expansion is not a good approximation 
for a strongly gravitating system, 
such as an extreme mass ratio binary system, 
an extrapolation is believed to give a good approximation of the result. 
In the perturbation limit, we have 
$\dot{\cal E}^{a(0,0)} \sim O((m/L)v^5)$, 
where $v$ is a typical velocity of the system, 
and $v^2$ is estimated to be less than $0.1$ 
during the inspiral stage of the binary. 
We have $((d/d\lambda)\lambda^b,(d/d\lambda)C^c)\sim O((m/L)v^5(t/L))$, 
and therefore the assumption of adiabaticity holds 
for the post-Newtonian radiation reaction time scale, 
$t<T_{PN.rad}\sim O(Lv^{-5}(m/L)^{-1})$, 
over which the adiabatic linear metric perturbation could be 
a solution of the Einstein equation to an accuracy of $O(m/L)$. 

As in the previous section, 
we suppose $m/L=10^{-5}$ with a dynamical time scale $L=100s$ 
for a possible target of the LISA project, 
and we have $T_{PN.rad}$ on the order of several years. 
Hence, we believe that 
the adiabatic linear metric perturbation in the radiation reaction gauge 
should give us a reliable prediction of gravitational waveforms 
for the LISA project.

\section{Summary} \label{sec:sum}

In this paper, we have considered the so-called self-force problem 
in the calculation of gravitational waves 
from an extreme mass ratio binary. 
A strategy previously employed for this problem was that, 
because we already have a method 
to calculate gravitational waves 
for a given stress-energy tensor, 
we calculate a self-force 
in order to derive the orbit of a particle. 
A general form of a self-force called the MiSaTaQuWa self-force, 
was derived in the usual metric perturbation scheme.\cite{match} 
We find that, in the usual metric perturbation scheme, 
the self-force is completely gauge dependent, 
and it can even be the case that the self-force vanishes. 

In Sec.\ref{sec:gg}, we argue that this is because 
the usual metric perturbation scheme is defined 
to allow an arbitrary gauge condition, 
and as a result, an inappropriate choice of the gauge condition 
would eventually lead to a nonperturbative evolution 
of the metric from an energetic point of view. 
Such a consequence cannot be avoided 
in the usual metric perturbation scheme, 
because the scheme requires 
that the linear metric perturbation must be induced by a background geodesic, 
and therefore the metric perturbation cannot be tracked 
with the orbital evolution caused by the self-force. 
Although the nonperturbative evolution of the metric 
could be calculated using numerical relativity, 
it would require a very large computational cost, 
and for this reason, we consider it advantageous 
to modify the metric perturbation scheme 
with a constraint of a gauge condition for the self-force. 
In Sec.\ref{sec:adi}, we propose an adiabatic metric perturbation scheme. 
In this new metric perturbation scheme, 
the so-called adiabatic approximation is applied to 
both the metric perturbation and the orbit. 
As a result, the metric perturbation is derived 
together with the orbital evolution caused by the gravitational self-force. 
We find that the new metric perturbation scheme is valid 
over the radiation reaction time scale, 
whereas the usual metric perturbation scheme is valid 
over only the dephasing time scale, 
which is shorter than the radiation reaction time scale 
by the square root of the mass ratio $(\sim 10^{-2.5})$. 

The adiabatic approximation has been successfully 
applied to linear metric perturbations, 
but it is not a simple problem to apply it 
to post-adiabatic metric perturbations 
because the linearized Einstein operator is 
algebraically divergence free, 
as we discussed briefly in Sec.\ref{sec:adi}. 
Although it is valid for sufficiently small 
$(d/d\lambda)\gamma$ and $(d^2/d\lambda^2)\gamma$, 
this may pose a question regarding the convergence of the adiabatic extension. 

For the self-force problem, there are two approaches: 
One employs the self-force calculation, 
which involves an explicit regularization calculation,\cite{self} 
and the other employs the radiation reaction formula,\cite{RR} 
which is a nontrivial extension of the balance formula.\cite{PT} 
In this paper, we have defined the adiabatic metric perturbation scheme 
using the radiation reaction formula. 
One may ask whether an adiabatic extension is possible 
for the self-force calculation. 
We believe that this would be nontrivial, because, for a technical reason, 
the self-force calculation cannot be easily carried out 
with a linear metric perturbation taking the form (\ref{eq:lm}). 
When the background is a Schwarzschild black hole, 
one may use the Zerilli-Regge-Wheeler formalism\cite{ZRW} 
to calculate a metric perturbation. 
In this case, one can transform the metric perturbation 
into the form (\ref{eq:lm}). 
However, when the background is a Kerr black hole, 
a method to calculate an inhomogeneous metric perturbation 
has only been proposed\cite{am} 
and it has not yet been implemented numerically. 
It is not clear that 
the metric perturbation in this method can be transformed 
into the form (\ref{eq:lm}). 

Because all we need for the orbital evolution in the end is 
a gauge invariant quantity $\dot{\cal E}^{a(0,0)}$, 
one may consider the possibility of calculating it 
by taking a long time average 
of the result of the self-force calculation,\cite{self} 
assuming that the self-force calculation is implemented 
under an appropriate gauge condition. 
Although this may sound a plausible approach, 
the self-force averaged over a long time $T$ becomes 
\begin{eqnarray}
\left<{d\over d\lambda}{\cal E}^a\right>_T &\sim& 
\sum \dot{\cal E}^{a(n_r,n_\theta)} \,, 
\end{eqnarray}
where the sum is taken over $n_r$ and $n_\theta$ for which 
$|n_r\tilde\Omega_r+n_\theta\tilde\Omega_\theta|<1/T$ 
(see (\ref{eq:rrf})). 
Thus, there is a case in which 
this approach fails to provide $\dot{\cal E}^{a(0,0)}$. 

By contrast, the radiation reaction formula\cite{RR} 
requires only a homogeneous metric perturbation 
and a conventional method for this calculation is known.\cite{chrz} 
Because the necessary calculation is 
only a slight extension of the balance formula, 
we consider the numerical technique for this calculation 
to be established. 
A homogeneous metric perturbation can be derived 
by operating with a tensor differential operator 
on the homogeneous Teukolsky function,\cite{chrz} 
and an analytic method for the homogeneous Teukolsky function 
is known.\cite{mano} 
Because of its fast convergence, 
this analytic method gives a more accurate result 
in a more efficient way.\cite{mano} 

In Sec.\ref{sec:gauge}, we introduced the radiation reaction gauge, 
in which the self-force is expressed exactly 
by the radiation reaction formula (\ref{eq:sfrr}). 
We recall that the energetic argument given in Sec.\ref{sec:gg} 
shows that the metric could be nonperturbative 
with an inappropriate choice of the gauge condition for the self-force. 
Because the usual metric perturbation scheme allows 
an arbitrary gauge condition, 
it is valid only over a short time scale. 
Contrastingly, the validity of the adiabatic metric perturbation 
depends on the gauge condition, 
because it solves the metric perturbation and the orbit at the same time. 
We find that the adiabatic metric perturbation 
under the radiation reaction gauge condition 
is valid over a longer time scale 
than that under a general gauge condition 
for the radiation reaction formula (\ref{eq:lm}). 
We also note that the evaluation of the validity 
might not be correct for a highly eccentric orbit. 

It is also interesting that 
a previous study using the balance formula\cite{PT0,PT1} 
uses the equation (\ref{eq:sfrr}) 
by the assumption of the adiabatic approximation. 
In this approach, the adiabatic approximation holds 
if the orbital period is sufficiently 
shorter than the dephasing time. 
However, the analysis in Sec.\ref{sec:gauge} shows 
that one can consistently eliminate 
the oscillating part of the self-force through the gauge, 
as long as it is small. 
Hence, the condition for (\ref{eq:sfrr}) must be 
the same for the adiabatic metric perturbation scheme.

\section*{Acknowledgements}

YM thanks Prof. Eric Poisson for fruitful discussions. 
YM was supported by NASA grant NAG5-12834 
and NASA-ATP grant NNG04GK98G from CalTech, 
where the idea of the adiabatic metric perturbation was formulated. 
YM was also supported by the National Science Foundation 
under Grant number NSF-PHY-0140326(UTB) 
and by NASA-URC-Brownsville 
when YM was at the University of Texas, Brownsville, 
where the idea of the radiation reaction gauge was formulated. 

\appendix

\section{Geodesic} \label{app:geo} 

We consider inspiralling geodesics in a Kerr geometry.\footnote{
A rather sophisticated analysis using Hamilton-Jacobi formalism 
is given Ref.\citen{geo}. 
Here we summarize another derivation given in Ref.\citen{RR} 
because it is convenient for the analysis given in this paper.} 
A geodesic equation is a set of 2nd order differential equations. 
Therefore, geodesics are characterized by $6$ integral constants. 
$3$ of them are constants of motion, 
the orbital energy ($E$), the $z$-component of the angular momentum ($L$) 
and the Carter constant ($C$). 
In order to distinguish the rest of the constants, 
we call $E$, $L$ and $C$ the primary constants 
and denote them by ${\cal E}^a,\,a=\{E,L,C\}$ when it is appropriate. 
Using the primary constants, the geodesic equation becomes 
\begin{eqnarray}
\left({dr \over d\lambda}\right)^2 
&=& (E(r^2+a^2)-aL)^2 -(r^2+(L-aE)^2+C)\Delta 
\,, \label{eq:g_r} \\ 
\left({d\theta\over d\lambda}\right)^2 
&=& C-a^2(1-E^2)\cos^2\theta -L^2\cot^2\theta
\,, \label{eq:g_the} \\ 
{dt \over d\lambda}
&=& -a(aE\sin^2\theta-L) +(r^2+a^2)(E(r^2+a^2)-aL)/\Delta
\,, \label{eq:g_t} \\ 
{d\phi \over d\lambda}
&=& -(aE-L\cosec^2\theta) +(a/\Delta)(E(r^2+a^2)-aL)
\label{eq:g_p} \,, \\ 
\Delta &=& r^2-2Mr+a^2 
\,. 
\end{eqnarray} 
Here we use $\lambda$ as an orbital parameter 
which has a one-to-one relation with the proper time, $\tau$, 
given by $\tau = \int d\lambda (r^2+a^2\cos^2\theta)^2$. 

For inspiraling geodesics, $r$/$\theta$-motion is bounded 
and the solutions of (\ref{eq:g_r}) and (\ref{eq:g_the}) 
can be expanded in discrete Fourier series as 
\begin{eqnarray}
z^b &=& \sum_n Z^{b(n)} \exp[in\chi_b] \,, \quad 
\chi_b \,=\, 2\pi\tilde\Omega_b(\lambda+\lambda^b)
\,, \label{eq:geo_r} 
\end{eqnarray}
where $b=\{r,\theta\}$, with $r=z^r$ and $\theta=z^\theta$, and 
$Z^{b(n)}$ and $\tilde\Omega_b$ are functions of ${\cal E}^a$. 
Here we have two integral constants $\lambda^b$, 
but, since we can freely choose the zero point of $\lambda$, 
$\lambda^r-\lambda^\theta$ alone specifies the geodesic. 
(\ref{eq:g_t}) and (\ref{eq:g_p}) are easily integrated, and we obtain 
\begin{eqnarray}
z^c &=& \kappa_c+\sum_{b,n}Z_b^{c(n)}\exp[in\chi_b] \,, \quad 
\kappa_c \,=\, <\dot Z^c>\lambda+C^c 
\,, \label{eq:geo_t} 
\end{eqnarray}
where $c=\{t,\phi\}$ with $t=z^t$ and $\phi=z^\phi$, 
and $\left<\dot Z^c\right>$ and $Z_b^{c(n)}$ are functions of ${\cal E}^a$. 
Here we have two integral constants $C^c$
which specify the geodesic. 
We call $\lambda^b$ and $C^c$ 
the secondary constants of the geodesic. 

Because of the periodicity, one can freely add 
$1/\tilde\Omega_b$ to $\lambda^b$. 
In general, the ratio of $\tilde\Omega_r$ to $\tilde\Omega_\theta$ 
is irrational, 
and one can take $\lambda^r-\lambda^\theta$ 
as small as possible using this freedom. 
The constants $C^c$ can be set zero 
by using the $t/\phi$-translation symmetry of the Kerr geometry.

\section{Self-Force} \label{app:sf} 

We argue that a metric perturbation defined by (\ref{eq:lm}) 
yields a general form of the self-force given in Ref.\citen{RR}. 
In Ref.\citen{RR}, 
a general form of the self-force (\ref{eq:rrf}) is derived, 
using a symmetry property of the Kerr geometry 
and a symmetry property of a geodesic. 
Because the self-force is induced by a geodesic, 
it is a function of the geodesic parameter $\lambda$ 
and the geodesic constants $\gamma =\{{\cal E}^a,\lambda^b,C^c\}$ 
(see Appendix \ref{app:geo} for definitions). 
Employing the symmetry of the Kerr geometry, 
Ref.\citen{RR} finds that the self-force can be independent of $C^c$. 
Then by a symmetry property of a geodesic, 
the self-force can be expanded in a discrete Fourier series 
of terms of the form 
$\exp[n_r(\lambda-\lambda^r)+n_\theta(\lambda-\lambda^\theta)]$. 
In this appendix, we demonstrate that 
the gauge condition for the metric perturbation (\ref{eq:lm}) 
actually respects the symmetry of the Kerr geometry 
by showing that the $C^c$ derivative of the self-force vanishes. 

We assume that the self-force 
acting on the orbital energy, angular momentum and Carter constant 
can be derived using a regularization calculation\cite{self} as 
\begin{eqnarray}
{d\over d\lambda}{\cal E}^a(\lambda) &=& 
\lim_{x\to z(\lambda;\gamma)}F^a[h(x;\gamma)-h^{sing}(x;\gamma)] \,, 
\end{eqnarray}
where $F^a[h]$ is a tensor derivative operator 
acting on the metric perturbation $h_{\mu\nu}$. 
$x^\mu$ and $z^\mu(\lambda;\gamma)$ are 
a field point and an orbital point of the geodesic $\gamma$ respectively. 
$h_{\mu\nu}(x;\gamma)$ is defined by (\ref{eq:lm}). 
$h^{sing}_{\mu\nu}(x;\gamma)$ is called 
the ``singular part'' of the metric perturbation, 
and it is defined in Ref.\citen{match} under the harmonic gauge condition. 

Using the temporal and rotational Killing vectors 
$\xi_{(c)}^\mu ,\,c=\{t,\phi\}$, 
the symmetry property of the Kerr geometry is defined 
in terms of the Lie derivative along the Killing vectors as 
\begin{eqnarray}
{\cal L}_{\xi_{(c)}}g^{kerr}_{\mu\nu}=0 \,. 
\end{eqnarray}
The property (\ref{eq:gm}) of the full Green function 
for a metric perturbation is expressed as 
\begin{eqnarray}
({\cal L}_{\xi_{(c)}}+{\cal L'}_{\xi_{(c)}})
G_{\mu\nu\,\alpha\beta}(x,z) =0 \,, \label{eq:fgg}
\end{eqnarray}
where ${\cal L}_{\xi_{(c)}}$ acts on $x$ and the indices $\mu$ and $\nu$, 
and ${\cal L'}_{\xi_{(c)}}$ acts on $z$ and the indices $\alpha$ and $\beta$.
The property (\ref{eq:gm}) is consistent 
with the linearized Einstein operator $G^{(1)}_{\mu\nu}$ 
since it commutes with the Killing vectors, i.e. 
\begin{eqnarray}
[{\cal L}_{\xi_{(c)}},G^{(1)}_{\mu\nu}] &=& 0 \,. 
\end{eqnarray}

It is possible to derive $h^{sing}_{\mu\nu}(x;\gamma)$ 
using the Green method with a singular Green function, 
and an explicit form of the singular Green function 
in the harmonic gauge condition is constructed 
using a bi-tensor expansion in Ref.\citen{match}. 
Because bi-tensors are defined in terms of background geometric quantities, 
the singular Green function in the harmonic gauge condition 
can also satisfy the relation 
\begin{eqnarray}
({\cal L}_{\xi_{(c)}}+{\cal L'}_{\xi_{(c)}})
G^{sing.}_{\mu\nu\,\alpha\beta}(x,z) =0 \,. \label{eq:sgg}
\end{eqnarray}
As in Ref.\citen{sfg}, we assume 
that a gauge transformation acts 
only on the full Green function.\footnote{
Strictly speaking, there are two types of gauge transformations 
for the MiSaTaQuWa self-force.\cite{match} 
One is for the metric of the far-zone expansion 
and is discussed in Ref.\citen{sfg}. 
The other is for the metric of the near-zone expansion. 
The metric of the near-zone expansion is defined 
as the sum of the black hole metric as a point particle 
and its perturbation. 
In Ref.\citen{match}, we choose the gauge condition for which 
the particle is located at $r=0$ of the Schwarzschild coordinates 
when the $L=1$ mode of the black hole perturbation vanishes. 
With this condition, the singular part is uniquely determined.} 
Then, the singular Green function has 
the property (\ref{eq:sgg}) in general. 

Because the singular behavior of the full Green function 
cancels with that of the singular Green function, 
one can define a regular Green function by 
\begin{eqnarray}
G^{reg}_{\mu\nu\,\alpha\beta}(x,z) &=& 
G_{\mu\nu\,\alpha\beta}(x,z) 
-G^{sing}_{\mu\nu\,\alpha\beta}(x,z) \,, 
\end{eqnarray}
and the self-force is simply derived as 
\begin{eqnarray}
{d\over d\lambda}{\cal E}^a(\lambda) &=& F^a[h^{reg}](z(\lambda);\gamma) 
\label{eq:rsf} \,, \\ 
h^{reg}_{\mu\nu}(x;\gamma) &=& \int dz^4 
G^{reg}_{\mu\nu\,\alpha\beta}(x,z) T^{\alpha\beta}(z;\gamma) \,, 
\end{eqnarray}
where $T^{\alpha\beta}$ is 
the usual stress-energy tensor of a point particle 
moving along the geodesic $\gamma$. 

Using (\ref{eq:fgg}) and (\ref{eq:sgg}), and integrating by parts, 
we have 
\begin{eqnarray}
{\cal L}_{\xi_{(c)}}h^{reg}_{\mu\nu}(x;\gamma) &=& \int dz^4 
G^{reg}_{\mu\nu\,\alpha\beta}(x,z) 
{\cal L'}_{\xi_{(c)}}T^{\alpha\beta}(z;\gamma) \,. 
\end{eqnarray}
From the result of Appendix \ref{app:geo}, we find 
\begin{eqnarray}
{\cal L'}_{\xi_{(c)}}T^{\alpha\beta}(z;\gamma) &=& 
-{\partial \over \partial C^c}T^{\alpha\beta}(z;\gamma) \,. 
\end{eqnarray}
Since the $C^c$ derivative of the self-force (\ref{eq:rsf}) 
acts on $z(\lambda)$ and $\gamma$ as 
\begin{eqnarray}
{\partial\over\partial C^c}F^a[h^{reg}](z(\lambda);\gamma) &=& 
\left\{{\cal L}_{\xi_{(c)}}F^a[h^{reg}](x;\gamma)
+{\partial\over\partial C^c}F^a[h^{reg}](x;\gamma)
\right\}_{x=z(\lambda)} \nonumber \\ 
&=& 0 \,, 
\end{eqnarray}
we find that the self-force is independent of $C^c$. 

\section{Evolution Equations} \label{app:orb}

The evolution equations for the primary ``constants'' 
are given by the self-force (\ref{eq:rrf}). 
In this appendix, we derive the evolution equations 
for the secondary ``constants'' $(\lambda^b,C^c)$. 
For convenience, we define the orbit as 
\begin{eqnarray}
z^b &=& \sum_n Z^{b(n)} \exp[in\tilde\chi_b] \,, \quad
z^c \,=\, \tilde\kappa_c+\sum_{b,n}Z_b^{c(n)}\exp[in\tilde\chi_b] \,,  
\end{eqnarray}
where $Z^{b(n)}$ and $Z_b^{c(n)}$ are the functions of ${\cal E}^a$ 
given in (\ref{eq:geo_r}) and (\ref{eq:geo_t}), 
and consider the evolution equations for $(\tilde\chi_b,\tilde\kappa_c)$. 
We can derive the evolution for $(\lambda^b,C^c)$ 
by using $\tilde\chi_b=2\pi\tilde\Omega_b(\lambda+\lambda^b)$ 
and $\tilde\kappa_c=<\dot Z^c>\lambda+C^c$ 
(see (\ref{eq:geo_r}) and (\ref{eq:geo_t})). 

Using $(\tilde\chi_b,\tilde\kappa_c)$, 
the self-force (\ref{eq:rrf}) is written as 
\begin{eqnarray}
{d\over d\lambda}{\cal E}^a &=& \tilde F^a 
\,=\, \sum_{n_r,n_\theta} \dot{\cal E}^{a(n_r,n_\theta)}({\cal E}^a)
\exp[in_r\tilde\chi_r+in_\theta\tilde\chi_\theta] 
\,. \label{eq:ev_e}
\end{eqnarray}
It is important to note that 
(\ref{eq:g_r})-(\ref{eq:g_p}) hold for non-geodesic orbits, 
because they are derived from the definitions of $(E,L,C)$, 
whether $(E,L,C)$ are constant or not. 
Together with (\ref{eq:ev_e}), 
(\ref{eq:g_r}) and (\ref{eq:g_the}) give 
the evolution equations for $\tilde\chi_b$, 
and we obtain the evolution equations for $\tilde\kappa_c$ 
from (\ref{eq:g_t}) and (\ref{eq:g_p}). 

We rewrite (\ref{eq:g_r}) and (\ref{eq:g_the}) as 
\begin{eqnarray}
\left({dz^b\over d\lambda}\right)^2 &=& Z^b({\cal E},z^b)
\,. \label{eq:g_r_the}
\end{eqnarray}
For convenience, we define 
\begin{eqnarray}
\delta_\lambda z^b &=& \sum_n
in2\pi\tilde\Omega_b Z^{b(n)} \exp[in\tilde\chi_b] \,, \quad
\delta_a z^b \,=\, \sum_n
Z^{b(n)}_{,a} \exp[in\tilde\chi_b] \,,
\end{eqnarray}
where $_{,a}$ represents the derivative with respect to ${\cal E}^a$.
Then, we have
\begin{eqnarray}
{dz^b\over d\lambda} &=&
{\delta_\lambda z^b \over 2\pi\tilde\Omega_b}
\left({d\tilde\chi_b\over d\lambda}\right)
+\sum_a(\delta_a z^b)\tilde F^a \,,
\end{eqnarray}
and, from (\ref{eq:g_r_the}), we obtain 
the evolution equations for $\tilde\chi_b$ as
\begin{eqnarray}
{d\tilde\chi_b\over d\lambda} &=& 2\pi\left(1-
{\sum_a\delta_a z^b\tilde F^a\over \delta_\lambda z^b}
\right)\tilde\Omega_b \,, \label{eq:ev_c}
\end{eqnarray}
where we have used $(\delta_\lambda z^b)^2 = Z^b({\cal E},z^b)$.

We next consider the $t/\phi$-motion.
We rewrite (\ref{eq:g_t}) and (\ref{eq:g_p}) as
\begin{eqnarray}
{dz^c\over d\lambda} &=& \sum_b Z_b^c({\cal E},z^b)
\,. \label{eq:g_t_p}
\end{eqnarray}
For convenience, we define 
\begin{eqnarray}
\delta_b z^c &=& \sum_n
in2\pi\tilde\Omega_b Z_b^{c(n)} \exp[in\tilde\chi_b] \,, \quad 
\delta_a z^c \,=\, \sum_{b,n} 
Z_{b\,,a}^{c(n)} \exp[in\tilde\chi_b] \,. 
\end{eqnarray}
Then, we have 
\begin{eqnarray}
{dz^c\over d\lambda} &=& {d\tilde\kappa_c\over d\lambda} 
+\sum_b \left({d\tilde\chi_b\over d\lambda}\right)
{\delta_b z^c \over 2\pi\tilde\Omega_b}
+\sum_a (\delta_a z^c)\tilde F^a \,, 
\end{eqnarray}
and, from (\ref{eq:g_t_p}), we obtain 
the evolution equations for $\tilde\kappa_c$ as 
\begin{eqnarray}
{d\tilde\kappa_c\over d\lambda} &=&
\left<\dot Z^c\right> 
+\sum_a \left(-\delta_a z^c
+\sum_b {\delta_a z^b \delta_b z^c\over \delta_\lambda z^b}\right)
\tilde F^a \,, \label{eq:ev_k} 
\end{eqnarray}
where we have used 
$\left<\dot Z^c\right>+\sum_b \delta_b z^c = \sum_b Z_b^c({\cal E},z^b)$. 

In summary, we obtain the evolution equations for $(\lambda^b,C^c)$ as 
\begin{eqnarray}
{d\lambda^b \over d\lambda} &=& -\sum_a
\left({\delta_a z^b\over \delta_\lambda z^b}
+{\tilde\Omega_{b,a}\over\tilde\Omega_b}(\lambda+\lambda_b)\right)
\tilde F^a \,, \label{eq:orb_l} \\
{dC^c \over d\lambda} &=& -\sum_a
\left(\delta_a z^c
-\sum_b {\delta_a z^b \delta_b z^c\over \delta_\lambda z^b}
+\left<\dot Z^c\right>_{,a}\lambda\right)
\tilde F^a \,. \label{eq:orb_c}
\end{eqnarray}
It is notable that, 
although we have the freedom 
to choose the zero point for $\lambda$
and to add $1/\tilde\Omega_b$ to $\lambda^b$, 
the last terms of (\ref{eq:orb_l}) and (\ref{eq:orb_c}) 
eventually grow linearly in $\lambda$. 
Therefore, $((d/d\lambda)\lambda^b,(d/d\lambda)C^c)$ 
are estimated to be $O((m/L)(t/L))$, 
and similarly $((d^2/d\lambda^2)\lambda^b,(d^2/d\lambda^2)C^c)$ 
behaves as $O((m/L)(t/L))$, 

\section{Orbital Evolution: Circular or Equatorial Case} 
\label{app:toy}

We consider integration of the orbital equations 
(\ref{eq:ev_e}), (\ref{eq:ev_c}) and (\ref{eq:ev_k}) 
in order to understand the properties of orbits. 
Here we consider only 
either an eccentric orbit in the equatorial plane ($\theta=\pi/2$)
or a circular orbit in an inclined precessing plane ($r=const$). 
In this case, it is sufficient
to consider just one of $\tilde\chi_b$ (or $\lambda_b$).
We also use the fact that
the primary ``constants'' can be treated in a perturbative manner
over the radiation reaction time scale.

With an appropriate scaling, the evolution equations 
(\ref{eq:ev_e}), (\ref{eq:ev_c}) and (\ref{eq:ev_k}) 
can be reduced to 
\begin{eqnarray}
{dx\over dt} &=& A +O(\mu^2) \,, \quad
A \,=\, \sum_n a_n e^{in y} \,, \label{eq:dx} \\
{dy\over dt} &=& 1+x+B +O(\mu^2) \,,\quad
B \,=\, \sum_n b_n e^{in y} \,, \label{eq:dy} \\
{dz\over dt} &=& 1+x+C +O(\mu^2) \,, \quad
C \,=\, \sum_n c_n e^{in y} \,, \label{eq:dz}
\end{eqnarray}
where $(x,y,z;t)$ corresponds to 
$({\cal E}^a,\tilde\chi^b,\tilde\kappa^c;\lambda)$. 
Here, $\mu=m/L<1$ is used as an index of the approximation. 
Thus, $(a_n,b_n,c_n)\sim O(\mu)$. 
The $O(\mu^2)$ terms of (\ref{eq:dx})--(\ref{eq:dz}) come 
from the nonlinear metric perturbation. 

For simplicity, we assume the initial values 
$x=y=z=0$ at $t=0$.
From (\ref{eq:dx}) and (\ref{eq:dy}), we have
\begin{eqnarray}
{dx\over dy} &=& A +O(\mu^2) \,,
\end{eqnarray}
and therefore we have
\begin{eqnarray}
x &=& A^{[1]} +O(\mu^2 y) \,, \label{eq:xx}
\end{eqnarray}
where
\begin{eqnarray}
A^{[1]} &=& a_0 y +\sum_n a^{[1]}_n e^{in y} \,, \nonumber \\ 
a^{[1]}_0 &=& -\sum_{n\not =0}a^{[1]}_n \,, \quad
a^{[1]}_n \,=\, {1\over in}a_n \,. \quad(n\not =0) 
\end{eqnarray}
From (\ref{eq:dz}) and (\ref{eq:dy}), we have
\begin{eqnarray}
{dz\over dy} &=& 1-B+C +O(\mu^2 y) \,,
\end{eqnarray}
and therefore we have
\begin{eqnarray}
z &=& y -B^{[1]} +C^{[1]} +O(\mu^2 y^2) \,, \label{eq:zz}
\end{eqnarray}
where
\begin{eqnarray}
B^{[1]} &=& b_0 y +\sum_n b^{[1]}_n e^{in y} \,, \nonumber \\ 
b^{[1]}_0 &=& -\sum_{n\not =0}b^{[1]}_n \,, \quad
b^{[1]}_n \,=\, {1\over in}b_n \,, \quad(n\not =0) \\
C^{[1]} &=& c_0 y +\sum_n c^{[1]}_n e^{in y} \,, \nonumber \\ 
c^{[1]}_0 &=& -\sum_{n\not =0}c^{[1]}_n \,, \quad
c^{[1]}_n \,=\, {1\over in}c_n \,. \quad(n\not =0) 
\end{eqnarray}

(\ref{eq:xx}) and (\ref{eq:zz}) show that
the evolution of $x$ and $z$ with respect to $y$
is dominatly determined by $a_0,b_0$ and $c_0$
when $y>1$ and $y>\mu$, respectively,
and the effect of $a_n,b_n$ and $c_n$ with $n\not =0$
is only $O(\mu)$.
We also note that the effect of the self-force 
by a nonlinear metric perturbation 
becomes $O(1)$ when $y\sim\mu^{-1}$, 
and therefore the approximation loses validity. 

(\ref{eq:dy}) becomes
\begin{eqnarray}
(1-A^{[1]}-B){dy\over dt} &=& 1+O(\mu^2 t) \,,
\end{eqnarray}
and we can integrate to obtain 
\begin{eqnarray}
y-A^{[2]}-B^{[1]} &=& t+O(\mu^2 t^2) \label{eq:yy0} \,,
\end{eqnarray}
where we define $A^{[2]}$ and $B^{[1]}$ as
\begin{eqnarray}
A^{[2]} &=& a_0 y^2/2 +a^{[1]}_0 y +\sum_n a^{[2]}_n e^{in y} 
\,, \nonumber \\ 
a^{[2]}_n &=& {1\over in}a^{[1]}_n \,(n\not =0) \,, \quad
a^{[2]}_0 \,=\, -\sum_{n\not =0}a^{[2]}_n \,.
\end{eqnarray}
We can rewrite (\ref{eq:yy0}) as
\begin{eqnarray}
y &=& {1-a_0^{[1]}-b_0\over a_0}
\left[1-\sqrt{1-{2a_0\over (1-a_0^{[1]}-b_0)^2}(t+D)}\right]
+O(\mu^2 t^2) \,, \label{eq:yy}
\end{eqnarray}
where we define $D=\sum_n (a^{[2]}_n+b^{[1]}_n)e^{in y}$.

(\ref{eq:yy}) shows that
the evolution of $y$ with respect to $t$
is dominantly determined by $a_0,a^{[1]}_0$ and $b_0$
when $t>\mu$,
and the contribution from $a_n,b_n$ and $c_n$ with $n\not =0$
is smaller by a factor of $O(\mu/t)$ 
relative to the leading order.
The approximation becomes invalid when $t\sim\mu^{-1}$
due to the non-linear self-force.
Since $y\sim\mu^{-1}$ for $t\sim\mu^{-1}$,
the linear self-force can be used to obtain 
a prediction of the orbital evolution for the time satisfying $t<\mu^{-1}$.
One can further extend (\ref{eq:yy}) as
\begin{eqnarray}
y &=& {1\over 1-a_0^{[1]}-b_0}(t+D)
+{a_0\over 2(1-a_0^{[1]}-b_0)^3}(t+D)^2
+\cdots +O(\mu^2 t^2) \,, \\
&=& t +(a_0^{[1]}+b_0)t
+a_0 t^2/2 +a_0^2 t^3/2 +\cdots +O(\mu,\mu^2 t^2) \,.
\end{eqnarray}
Since $a_0^{[1]}$ depends on the choice of initial values, 
this shows that only $a_0$ is essentially important in (\ref{eq:dx}). 
This also suggests that (\ref{eq:orb}) 
is a qualitatively good approximation 
over the radiation reaction time scale, 
because $a_0 t^2/2>a_0^2 t^3/2$. 

\section{Gauge Transformation} \label{app:gau}

To evaluate (\ref{eq:ec_gau}),
we choose the gauge transformation vector as
\begin{eqnarray}
\xi^\alpha &=& \xi^E \eta^{E\,\alpha} +\xi^L \eta^{L\,\alpha}
+\xi^C \eta^{C\,\alpha}{}_\beta v^\beta \,.
\end{eqnarray}
Then, the transformation of the primary constants
defined in the perturbed spacetime becomes
\begin{eqnarray}
\delta_\xi{\cal E}^a &=& \sum_{b=E,L,C}\left(
A^a{}_b {d\over d\lambda}\xi^b +B^a{}_b \xi^b \right) \,,
\end{eqnarray}
where the $3\times 3$ matrices $A^a{}_b$ and $B^a{}_b$ are defined as
\begin{eqnarray}
A^a{}_b &=& \left(
\begin{array}{ccc}
-\eta^E_\alpha \eta^{E\,\alpha} &
-\eta^E_\alpha \eta^{L\,\alpha} &
-\eta^E_\alpha \eta^{C\,\alpha}{}_\beta v^\beta \\
-\eta^E_\alpha \eta^{L\,\alpha} &
-\eta^L_\alpha \eta^{L\,\alpha} &
-\eta^L_\alpha \eta^{C\,\alpha}{}_\beta v^\beta \\
-\eta^E_\alpha \eta^{C\,\alpha}{}_\beta v^\beta &
-\eta^L_\alpha \eta^{C\,\alpha}{}_\beta v^\beta &
-\eta^C_{\alpha\beta} \eta^{C\,\alpha}{}_\gamma
v^\beta v^\gamma
\end{array}
\right){d\lambda\over d\tau} \,, \\
B^a{}_b &=& \left(
\begin{array}{ccc}
0 &
v_\alpha [\eta^E,\eta^L]^\alpha &
\eta^C_{\alpha\beta}v^\alpha {D\over d\tau}\eta^{E\,\beta} \\
-v_\alpha [\eta^E,\eta^L]^\alpha &
0 &
\eta^C_{\alpha\beta}v^\alpha {D\over d\tau}\eta^{L\,\beta} \\
-\eta^C_{\alpha\beta}v^\alpha {D\over d\tau}\eta^{E\,\beta} &
-\eta^C_{\alpha\beta}v^\alpha {D\over d\tau}\eta^{L\,\beta} &
0
\end{array}
\right) \,.
\end{eqnarray}
For a general orbit, one can assume that
the symmetric matrix ${\bf A}$ has an inverse matrix, 
since the appropriate linear combinations of
$\eta^{E\,\alpha},\eta^{L\,\alpha}$ and $\eta^{C\,\alpha}{}_\beta v^\beta$
give three independent vectors.
When the orbit is equatorial (or quasi-circular),
the symmetric matrix could be singular. 
However, since it is sufficient
to consider $\{E,L\}$ (or $\{E,C\}$) in this case,
the relevant $2\times 2$ sub-matrix of ${\bf A}$
still have an inverse.


\end{document}